\documentclass[%
superscriptaddress,
 amsmath,amssymb,
 aps,
 twocolumn,
pra,
]{revtex4-2}

\usepackage{graphicx}
\usepackage{mathtools}
\usepackage{dcolumn}
\usepackage{bm}
\usepackage[colorlinks=true,
            linkcolor=red,
            urlcolor=blue,
            citecolor=blue]{hyperref}
\usepackage{graphicx} 
\usepackage{amsmath}
\usepackage{bm}
\usepackage{todonotes}
\usepackage[section]{placeins}
\usepackage{soul}

\newcommand{\NTT}{NTT Basic Research Laboratories, NTT Corporation, 3-1 Morinosato Wakamiya, Atsugi, Kanagawa, 243-0198, Japan}

\newcommand{\TQP}{NTT Research Center for Theoretical Quantum Physics, NTT Corporation, 3-1 Morinosato Wakamiya, Atsugi, Kanagawa, 243-0198, Japan}

\begin{document}

\preprint{APS/123-QED}

\title{Coherent response of inhomogeneously broadened and spatially localized emitter ensembles in waveguide QED
}


%
 
\author{L. Ruks}
\email{Lewis.ruks@ntt.com} 
\affiliation{\NTT}
\affiliation{\TQP}
\author{X. Xu}
\affiliation{\NTT}
\author{R. Ohta}
\affiliation{\NTT}
\author{W. J. Munro}
\affiliation{\TQP}
 
\author{V. M. Bastidas}
\affiliation{\NTT}
\affiliation{\TQP}

\date{\today}

\begin{abstract}
\noindent Spectrally and spatially varying ensembles of emitters embedded into waveguides are ever-present in both well-established and emerging technologies. If control of collective excitations can be attained, a plethora of coherent quantum dynamics and applications may be realized on-chip in the scalable paradigm of waveguide 
quantum electrodynamics (WQED).
Here, we investigate inhomogeneously broadened ensembles embedded with subwavelength spatial extent into waveguides employed as single effective and coherent emitters. We develop a method permitting the approximate analysis and simulation of such mesoscopic systems featuring many emitters, and show how collective resonances are observable within the waveguide transmission spectrum once their linewidth exceeds the inhomogeneous line. In particular, this  allows 
for near-unity and tailorable non-Lorentzian extinction of waveguide photons overcoming large inhomogeneous broadening present in current state-of-the-art. As a particular illustration possible in such existing experiments, we consider the classic emulation of the cavity QED (CQED) paradigm here using ensembles of rare-earth ions as coherent mirrors and qubits and demonstrate the possibility of strong coupling given existing restrictions on inhomogeneous broadening and ensemble spatial extent. This work introduces coherent ensemble dynamics in the solid-state to WQED and extends the realm to spectrally tailorable emitters.

\end{abstract}

\maketitle

%
\section{Introduction}
Ensembles of emitters in solid-state media are a valuable resource for shaping light and processing information as the matter component of hybrid optical platforms~\cite{Kurizki_2015}. Possible long individual coherence times combined with wide spectral bandwidth~\cite{serrano2022ultra,ohta2021rare} of the inhomogeneous line permit applications from quantum memories to atomic frequency combs~\cite{yasui2022creation,lago2021telecom}. When collectively addressed, ensembles enjoy large collective couplings to light and can be employed on the mesoscopic scale as single optical elements~\cite{ohta2021rare,zhong2017interfacing,zhu2011coherent,Amsss2011,Lei2023}. Specifically, ensembles embedded into waveguides benefit from well-established telecoms technologies~\cite{becker1999erbium,rinner2023erbium,Weiss2021,Mor2022} whilst allowing for an integrated and scalable optical platform~\cite{Sipahigil2016,Faraon2011}. Despite experimental demonstrations such as potential quantum memories~\cite{lago2021telecom,Choi2008,Julsgaard2013} and atomic frequency combs~\cite{yasui2022creation,Afzelius_2009}, on-chip operation in the framework of waveguide QED~\cite{sheremet2023waveguide} remains relatively unexplored theoretically in the regime when both spectral variation~\cite{kling2023characteristics,Song:21} and subwavelength finite spatial variation~\cite{Ruostekoski2016} are present. The latter feature -- recently feasible experimentally~\cite{Nandi:21,pak2022long,integrated_22} -- is particularly pertinent to the WQED paradigm. Here, ordered systems of point-like emitters enjoy unique dynamics featuring non-trivial excitation profiles~\cite{kim2021quantum} and many-body states~\cite{mirhosseini2019cavity,douglas2015quantum,See2019} enabling distinct functionality from CQED, with promise for photonic state generation~\cite{gonzalez2017efficient} and quantum simulation~\cite{douglas2015quantum}. An analogous realization of WQED taking into account unavoidable spatial disorder and inhomogeneous broadening  could open the door to a range of applications benefiting from the unique long coherence times and broadband nature of solid-state emitters, whilst exhibiting naturally scalability when compared with ensemble CQED platforms. {Beyond existing results on single-photon ensemble superradiance~\cite{Manassah_geometries}, it is also necessary to further understand the spectral response of disordered but spatially localized ensembles of finite extent and featuring inhomogeneous broadening in the context of current solid-state waveguide platforms. In particular, this includes cases where composite elements comprising multiple distinct ensembles are considered.}
\\
\indent In this work, we study ensembles of spatially localized waveguide-embedded emitters  featuring spectral inhomogeneity as candidates for effective and collectively enhanced optical elements in waveguides. We further investigate the joint conditions on spectral and positional inhomogeneity for collective coherence to emerge~\cite{Braggio2020} and dominantly establish the symmetrically excited polariton as an effective and coherent emitter excitation. To treat mescoscopic system sizes of $10^9$ emitters, we introduce a method to define collective spins via bins in both positional space and the frequency domain, enabling a 
reduction in computational cost and allowing us to demonstrate the formation of emitter coherence via observations of broad and near-unit extinction in the transmission statistics of photons through the waveguide. This method also retains the narrower collective resonances present within the familiar superradiant line~\cite{Manassah_geometries} and allows for an interpretation of loss of coherence as a coupling to these darker resonances within the ensemble as the spatial extent is increased. Interfacing of multiple and sufficiently spatially localized ensembles then suggests a realisation of the WQED paradigm~\cite{sheremet2023waveguide} with spectrally tailorable emitters. As a proof of principle we demonstrate the emulation of CQED -- including strong coupling -- amongst realistic ensembles of rare-earth ions in analogy with the single-emitter case~\cite{mirhosseini2019cavity,chang2012cavity} and demonstrate coherent operation for appreciable, but subwavelength, spatial ensemble extent within the reach of current state-of-the-art. 
We show how the emitter-density threshold required for strong coupling may be considerably reduced within existing experimental capabilities by shaping of the inhomogeneous line using spectral hole burning~\cite{moerner1988persistent}. Our results suggest that collective excitations in spatially localized solid-state ensembles can be exploited as effective emitters in near-term technologies within the paradigm of WQED, whilst augmenting this setting with spectral tailorability to explore new regimes of collective ensemble interaction beyond CQED. \\
\begin{figure}[t!]
\centering
\includegraphics[scale=0.65]{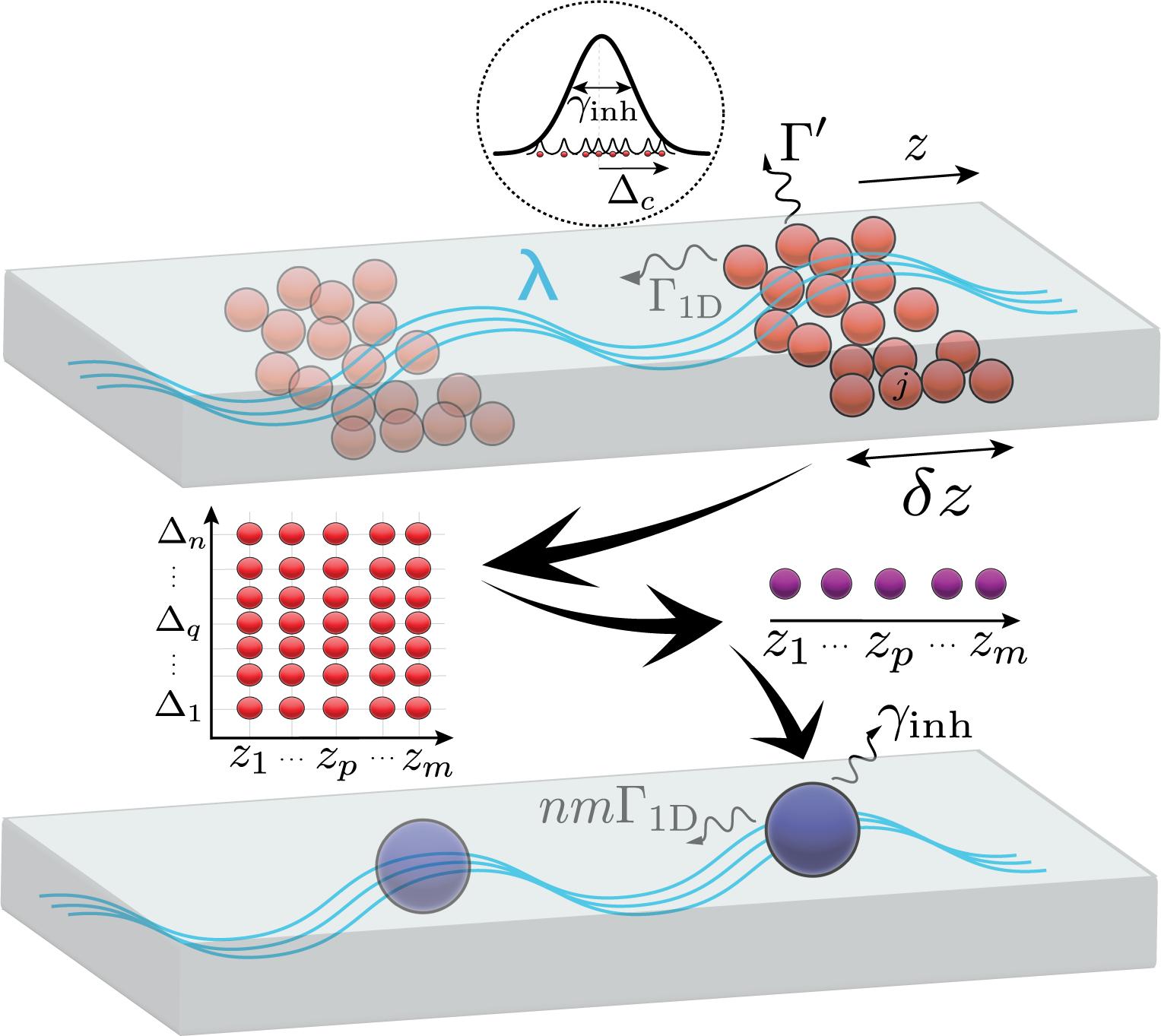}
    \caption{
    Binning and formation of a single effective emitter (large blue circle, bottom) for a localized ensemble of emitters (small red circles, top).}
\label{fig:schematic}
\end{figure}
\section{The model} In this work we consider $N$ two-level emitters that interact with a continuum of waveguide modes through the standard dipolar light-matter coupling. In addition to rare-earth ions considered here, the following analysis could be applied to systems of Doppler-broadened gases~\cite{ritter2018coupling}, quantum dots~\cite{norman2019review}, and NV centers~\cite{zhu2011coherent}. 
Assuming weak driving, the emitters in the rotating frame experience a distribution of detunings $\Delta_{j}$ with a full-width-at-half-maximum (FWHM) $\gamma_{\text{inh}}$. For weak emitter-field coupling, the field within the waveguide may be traced out to yield the Born-Markov master equation $	\dot{\hat{\rho}}_{\text{S}} = -\frac{\mathrm{i}}{\hbar}[\hat{H},\hat{\rho}_{\text{S}}] + \hat{\mathcal{L}}_{\textrm{coll}}[\hat{\rho}_{\text{S}}] + \hat{\mathcal{L}}_{\textrm{ind}}[\hat{\rho}_{\text{S}}]$ for emitter density matrix $\hat{\rho}_{S}$. The Hamiltonian and Linbladian generating system dynamics are given through the standard spin operators for the $j$-th emitter, $\hat{\sigma}_j^{\pm}$ and $\hat{\sigma}_j^{ee}=(\hat{\sigma}_j^z+\hat{1})/2$ (with $\{\hat{A},\hat{B}\}=\hat{A}\hat{B}+\hat{B}\hat{A}$ the anticommutator, while $\mathfrak{R}[w] = \text{Re}[w]$ and $\mathfrak{I}[w] = \text{Im}[w]$):
\begin{eqnarray}
\label{eq:master_eqn}
    \hat{H} &=&\hat{H}_{\text{em}}+ \hat{H}_{\text{drv}} +\frac{\hbar \Gamma_{\text{1D}}}{2}\sum_{j,k=1}^{N}\mathfrak{I}[G_{j,k}]\hat{\sigma}_{j}^{+}\hat{\sigma}_{k}^{-},
    \nonumber \\
    \hat{\mathcal{L}}_{\mathrm{coll}}[\hat{\rho}_{\text{S}}] &=& \frac{\Gamma_{\text{1D}}}{2}\sum_{j,k=1}^{N}\mathfrak{R}[G_{j,k}]\left(2\hat{\sigma}_{k}^{-}\hat{\rho}_{\text{S}}\hat{\sigma}_{j}^{+} - \{\hat{\sigma}_{j}^{+}\hat{\sigma}_{k}^{-},\hat{\rho}_{\text{S}}\} \right)
   \nonumber \\
    \hat{\mathcal{L}}_{\mathrm{ind}}[\hat{\rho}_{\text{S}}] &=& \frac{\Gamma'}{2}\sum_{j=1}^{N}\left(2\hat{\sigma}_{j}^{-}\hat{\rho}_{\text{S}}\hat{\sigma}_{j}^{+} - \{\hat{\sigma}_{j}^{+}\hat{\sigma}_{j}^{-},\hat{\rho}_{\text{S}}\}\right)
    .
\end{eqnarray}
Here, $\hat{H}_{\text{em}} = \hbar\sum_{j=1}^{N}\Delta_{j}\hat{\sigma}_{j}^{ee}$, and the drive reads $\hat{H}_{\text{drv}} = \hbar\sum_{j}\left(\Omega_{j}\hat{\sigma}^{+}_{j} + \Omega_{j}^{*}\sigma_{j}^{-}\right)$ for {$\Omega_{j} = \Omega(z_{j}),$} with an individual emitter decay with rate $\Gamma'$~\footnote{$\Gamma'$ may include photonic decay into unguided modes, or non-radiative decay into, e.g., material phonons.}. Crucially, the 
waveguide-mediated emitter-emitter interactions are subject to infinite-range interactions through the 1D propagator {$G_{j,k}=e^{\mathrm{i}\beta |z_{j}-z_{k}|}$}, with $z_j$ being the position of the $j$-th emitter along the waveguide. $\Gamma_{\text{1D}}$ is the single-emitter decay rate into the waveguide, and $\beta = 2\pi/\lambda$ is the wavenumber of the (assumed single) waveguided mode with wavelength $\lambda$.  {In the case that the waveguided mode field profile varies significantly over the emitter distribution in the transverse plane, it is sufficient to interpret $\Gamma_{\text{1D}}$ as an average of single-emitter decay rates into the waveguide (see Appendix \ref{app:transverse}) and continue to consider positions $z_{j}$ only. Inhomogeneous broadening deviating from that of ensembles in bulk media may also be assumed present in the distribution $\Delta_{j}$ }.  
\\
\section{Treating inhomogeneous broadening for single ensembles}
To address mesoscopic sizes of $N = 10^{9}$ and beyond in the general presence of inhomogeneity, we approximate the spatial-spectral density in the large-number limit as a decorrelated product of position and frequency densities such that each individual emitter lies in some designated frequency bin and position bin (Fig. ~\ref{fig:schematic}). 
{
Assuming $n$ frequency bins and $m$ positional bins with $nm=N$ and {$n,m\gg 1$}, we relabel each spin $j \to (p,q)$ 
such that $\hat{\sigma}^{ee}_{p,q}$ and $\hat{\sigma}_{p,q}^{\pm}$ correspond to the emitter in the $p$-th positional bin (at position $z_{p}$) and the $q$-th frequency bin (at detuning $\Delta_{q})$. The total emitter density in $(z,\Delta)$-space is determined by the distributions of $z_{p}$ and $\Delta_{q}.$
We consider the limit of low light intensity~\cite{Ruostekoski2016} for sufficiently small $\Omega$ where  $\langle\hat{\sigma}^{\alpha}_{j}\hat{\sigma}^{\beta}_{l}\rangle \approx \langle\hat{\sigma}^{\alpha}_{j}\rangle \langle\hat{\sigma}^{\beta}_{l} \rangle$ for $j\neq l$ and  $\langle \sigma_{j}^{ee}\rangle \approx 0$ for all $j$. This allows us to obtain the equations of motion for averages $\bullet \coloneqq 
\langle \hat{\bullet}\rangle$:
\begin{equation}
\dot{\sigma}_{p,q}^{-} = \mathrm{i}\Delta_{q}\sigma_{p,q}^{-} -\frac{\Gamma'}{2}\sigma_{p,q}^{-} -  \frac{\Gamma_{\text{1D}}}{2}\sum_{p',q'}G_{p,p'}\sigma^{-}_{p'q'} + i\Omega_{p},
\label{eq:linear-eq}
\end{equation}
where from Eq.~\eqref{eq:master_eqn} the only coefficients depending on $q$ are frequencies $\Delta_{q}$. As is physically expected, $\Delta_{q}$  is also assumed independent of position $z_{p}$ along the waveguide.
For each positional bin $p$ we define the symmetric lowering operator~\cite{diniz2011strongly} acting on the collective spin (purple emitter, Fig.~\ref{fig:schematic}) at $z_{p}$,
\begin{equation}
\hat{\mathcal{B}}_{p}^{-} = \frac{1}{\sqrt{n}}\sum_{q=1}^{n}\hat{\sigma}_{p,q}^{-}.
\end{equation}
In the steady-state we find the self-consistent linear-response relation for collective-spin coherences $\mathcal{B}^{-}_{p}$,
%
%
\begin{eqnarray}
\label{eq:linear_response}
    \mathcal{B}_{p}^{-} = i\sqrt{n}\gamma_{\text{inh}}^{-1}\chi(\Delta_{c})\left[ -  \frac{\sqrt{n}\Gamma_{\text{1D}}}{2}\sum_{p'=1}^{m}G_{p,p'}\mathcal{B}_{p'}^{-} + i\Omega_{p}\right],
\end{eqnarray}
whose number of degrees of freedom has been reduced from $N$ to $m$. Here the ensemble response function is defined $\chi(\Delta_{c}) = \gamma_{\text{inh}}\int \frac{d \Delta'\rho(\Delta')}{\Delta_{c} - \Delta' + \mathrm{i}\frac{\Gamma'}{2}}$ 
as the assumed continuum limit of $\frac{1}{n}\sum_{q}\frac{1}{\Delta_{c} - \Delta_{q} + i\frac{\Gamma'}{2}}$ for $n \gg 1$~\cite{wang2002}. Driving detuning $\Delta_{c}$ and $\Delta'$ entering the spectral density $\rho(\Delta')$ are defined as detunings from the inhomogeneous line mean. Interactions in \eqref{eq:linear_response} correspond to a non-Hermitian Hamiltonian
\begin{equation}
\label{eq:nh-hamiltonian}
\hat{H}_{\text{eff}}^{\text{nh}} =  -\mathrm{i}n\frac{\hbar\Gamma_{\text{1D}}}{2}\sum_{p,p'=1}^{m}G_{p,p'}\hat{\mathcal{B}}_{p}^{+}\hat{\mathcal{B}}_{p'}^{-}.
\end{equation}
%
Thus, the system \eqref{eq:linear_response} constitutes $m$ identical collective spins coupled to the waveguide at positions $z_{p}$ and featuring non-Lorentzian responses $\chi$. We use this representative system in the reduced state space of size $m$ to study mesocopic steady-state dynamics approximating that of \eqref{eq:linear-eq}, and justify its validity in Appendix \ref{app:binning}}

%
\textit{Transmission statistics.}
{
We here consider steady-state transmission through the waveguide-collective-spin system, defined through \eqref{eq:linear_response}. The transmission coefficient $t(\Delta_{c})$ describes the phase shift and attenuation of a waveguided incident coherent field with detuning $\Delta_{c}$ and $\Omega(z) = \Omega e^{i\beta z}$, where $\Omega$ may be arbitrary in the linear regime \eqref{eq:linear_response}. Finding transmission through the collective-spin system amounts to a substitution of the usual single-atom response $(\Delta_{c} + i\Gamma'/2)^{-1}$ by the collective spin response $\chi(\Delta_{c})/\gamma_{\text{inh}}$ in the standard expression for transmission through identical  Lorenztian atoms~\cite{sheremet2023waveguide,asenjo2017atom}:
\begin{equation}
\label{eq:transmission}
	t(\Delta_{c}) = \prod_{\mu=0}^{m-1}\left( \frac{\gamma_{\text{inh}}\chi^{-1}(\Delta_{c})}{\gamma_{\text{inh}}\chi^{-1}(\Delta_{c}) + \Lambda_{\mu}}\right).
\end{equation}
Here $\Lambda_{\mu} = \omega_{\mu} + \mathrm{i}\Gamma_{\mu}/2$ are the complex energy eigenvalues of $[\frac{in\Gamma_{\text{1D}}}{2}G_{p,p'}]_{pp'}$ that define collective excitations with decay rate $\Gamma_{\mu}$ and resonant frequencies shifted $\omega_{\mu}$ from the inhomogeneous line. Assuming the typical condition $\Gamma'/\gamma_{\text{inh}} \ll 1$ observed in solid-state ensembles, with $\rho$ symmetric, a large $|\Delta_{c}| \gg \gamma_{\text{inh}}$ expansion yields $ \gamma_{\text{inh}}\chi^{-1}(\Delta_{c}) \sim \Delta_{c} +\mathrm{i}\pi\Delta_{c}^2\rho(\Delta_{c}) 
 + i\Gamma'/2 + O(\gamma_{\text{inh}}^2/\Delta_{c}) $ at first order in $\Gamma'/\gamma_{\text{inh}}$~\cite{diniz2011strongly,zhong2017interfacing} . A consequence is that any collective resonance $\mu$ with $|\omega_{\mu}| \gtrapprox \gamma_{\text{inh}} \gg \Gamma'$ is observed in the transmission spectrum \eqref{eq:transmission} with an effective linewidth of approximately $(\Gamma' + \Gamma_{\mu} + 2\pi\Delta_{c}^2\rho(\Delta_{c}))/2$. This result, observed in the later section, is of practical interest as the open-waveguide analog of cavity protection~\cite{diniz2011strongly,zhong2017interfacing}: effects of broadening on far-shifted collective resonances can be mitigated by shaping the inhomogeneous line so that $\Delta_{c}^2\rho(\Delta_{c})\to 0$ for $\Delta_{c} \to \infty$}. 
%
\begin{figure}[t!]
\centering
\includegraphics[width=\linewidth]{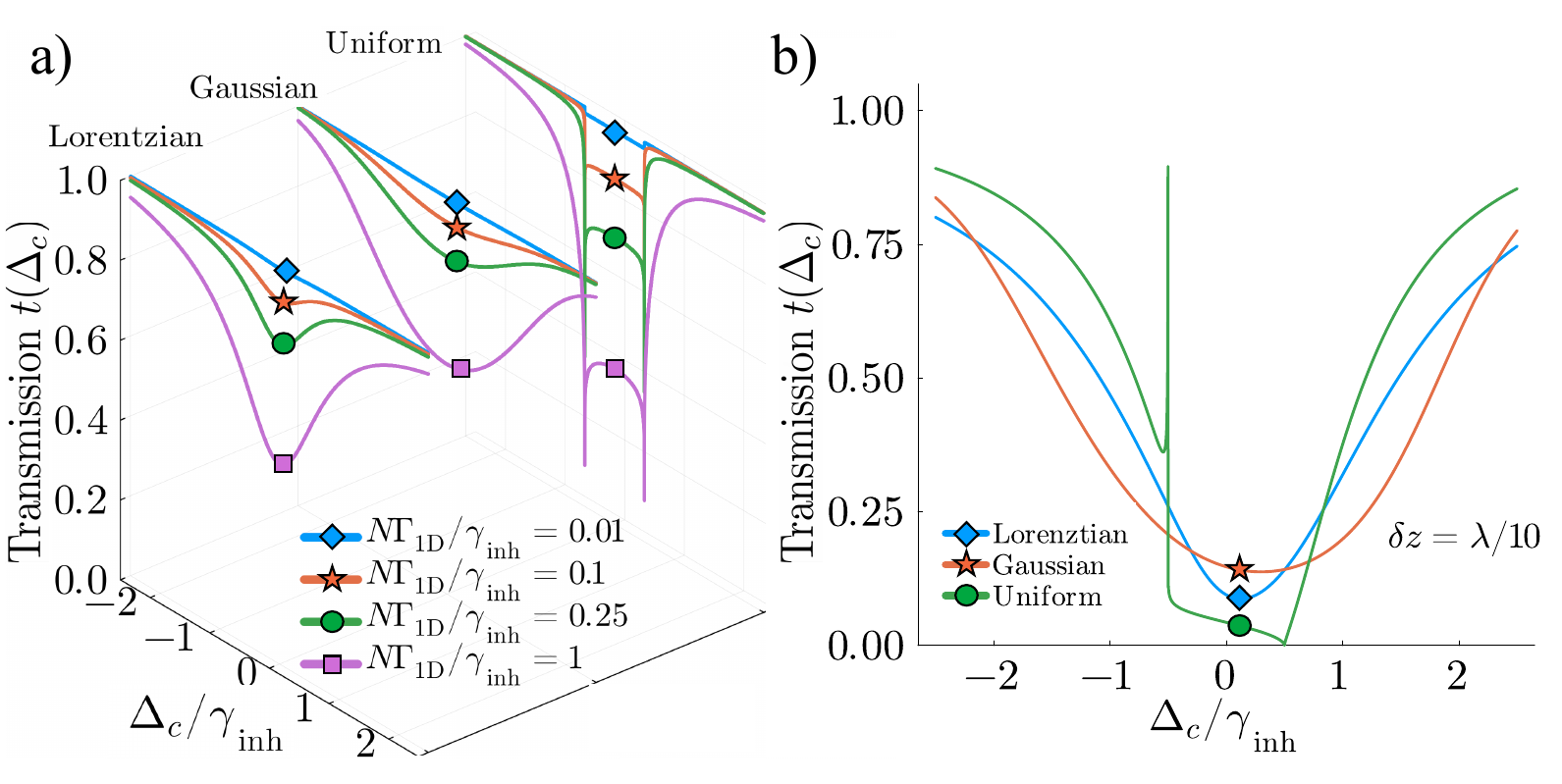}
    \caption{Single-ensemble transmission.
    (a) Transmission through an ensemble featuring no positional inhomogeneity as the collective waveguide coupling efficiency $N\Gamma_{\text{1D}}/\gamma_{\text{inh}}$ is increased for each of the three prototypical spectral distributions. {We take $\Gamma' = 10^{-6}\gamma_{\text{inh}}$}. (b) Transmission profile for an ensemble of $N=10^9$ emitters with $\gamma_{\text{inh}}/(2\pi)$ = 50GHz, $\Gamma_{\text{1D}}/(2\pi) = \Gamma'/(2\pi) = 100$Hz. 
    {We consider $m=10^3$ positional bins in Eq. \eqref{eq:linear_response}, for one realization of $z_{p} \sim U(0,\delta z)$}.}
\label{fig:single_ensemble}
\end{figure}
\\
\subsection{Coherent extinction}
Similarly to the case of single emitters~\cite{chang2012cavity,liao2016photon}, an inhomogeneous ensemble of emitters may act as a coherent mirror~\cite{Li2015,Nandi:21} for waveguided photons when the collective decay process exceeds the (effective) single-emitter linewidth. Here, the effective single-emitter linewidth is approximately set by the inhomogeneous linewidth. For a completely localized ensemble with identical $z_{p}$, the only non-zero eigenvalue is $\Lambda_{0} = \frac{in\Gamma_{\text{1D}}}{2},$ whilst $\Lambda_{\mu} = 0$ otherwise. As could be seen from Eq.~\eqref{eq:transmission}, the condition
{
\begin{eqnarray}
\label{eq:freq_inhomo_limit}
N\Gamma_{\text{1D}} \gg \gamma_{\text{inh}} \gg \Gamma',
\end{eqnarray}
}
then yields the appreciable, single broad resonance within the transmission in Fig.
\ref{fig:single_ensemble}(a) for the prototypical spectral distributions of FWHM $\gamma_{\text{inh}}$ defined in Appendix \ref{app:response}. 
This also coincides with the condition for the response in the right-hand side in Eq. \ref{eq:linear_response} to become appreciable, and establish significant coherence in the collective spin. When Eq. \eqref{eq:freq_inhomo_limit} holds, collective emission dominates and the observed extinction approximately corresponds to  reflection~\cite{chang2012cavity}. In addition, establishing of the resonance is accompanied by an appreciable phase shift of the transmitted photon~\cite{Domokos2002}. The onset of high quality reflectance can be further seen in Fig. \ref{fig:single_ensemble}(a) to be advanced by shaping the spectral distribution  using spectral hole burning. This effect can greatly relax density requirements for observing collective coherence, and holds up in the presence of appreciable subwavelength positional spread. 

\subsection{Inclusion of finite spatial extent}
Positional spread of emitters on the order $\delta z \lessapprox 0.1\lambda$ can be achieved in a variety of optical platforms~\cite{meng2018near,pak2022long,pfeiffer2014eleven}, and via ion-implantation specifically in the case of rare-earth ions~\cite{pak2022long}. In addition, well-below subwavelength confinement is available to microwave-based platforms~\cite{zhu2011coherent} and so from here we restrict our analytical analysis to the perturbative regime $\delta z \ll \lambda$. {Whilst in a spectrally homogeneous ensemble this single-photon superradiance condition is well-established~\cite{Manassah_geometries}, we here provide further analysis including the narrower resonances and in the presence of inhomogeneous broadening.} {From here we additionally assume a typical uniform distribution $z_{p} \sim U(0,\delta z),$ for positional bins}. 
In order to maintain coherent mirror-like operation, {we require the broad resonance observed in Fig. \ref{fig:single_ensemble}(a) to persist under the effect of small $\delta z > 0$, with effects limited to relatively narrow central transparency windows and an small overall shift of of the broad line~\cite{zhang2019theory,sheremet2023waveguide}. Driving around the broadest resonances excites slowly varying polarizations profiles over the length of the ensemble, which remain relatively unperturbed with respect to small positional fluctuations observed for large $m.$ As such, the eigenvalues $\Lambda_{\mu}$ with largest linewidth can be approximated by those obtained assuming uniform spatial separation $z_{p} = p\delta z/m$ for $p = 1,\ldots,m$.} The latter are derived in Appendix \ref{app:eigval_approx} to order $O(\nu^2)$ for $\nu = \beta \delta z \ll 1$,
\begin{eqnarray}
\label{eq:perturb_lines}
	&\Lambda_{0} = \frac{N\Gamma_{\text{1D}}}{2}\left(-\frac{\nu}{3} + \mathrm{i}\left[1 -\frac{4\nu^2}{45}\right]\right), \\
	&\Lambda_{\mu} = \frac{N\Gamma_{\text{1D}}}{2}\left(\frac{2\nu}{\mu^2\pi^2} + \mathrm{i}\frac{8\nu^2}{\mu^4\pi^4}\right) \ \ \ (1 \leq \mu \ll m),
\end{eqnarray}
{such that a small, finite $\delta z$ introduces $m-1$ narrow resonances within the center of the broad line~\cite{Vladimirova1998}. The $1/\mu^{4}$ scaling of linewidths means that resonances can be neglected in practice already for, say, $m \gg \mu \gtrapprox 10$ when compared to the scale of $\gamma_{\text{inh}}$. Retaining the first few broadest lines is then the reason for assuming $m \gg 1.$} The condition on linewidths, $\mathfrak{I}[\Lambda_{1}] \lessapprox \mathfrak{I}[\Lambda_{0}]$, to essentially maintain the single resonance is then obtained, \begin{equation}
\label{eq:position_inhomo_limit}
	\delta z \lessapprox 2\lambda/5,
\end{equation} {which is consistent with  the standard single-photon superradiance condition $\delta z \ll \lambda$.} Eq. \eqref{eq:position_inhomo_limit} is already well satisfied for $\delta z = 0.1\lambda$,  preserving the central broad line in Fig. \ref{fig:single_ensemble}(b). {In the joint presence of inhomogeneous broadening, we then see from Eq. \eqref{eq:transmission} that $\mathfrak{I}[\Lambda_{\mu}] \gg \gamma_{\text{inh}}$ is required more generally for the resonance $\mu$ of an ensemble to be visible in the transmission spectrum, so that increasing spatial extent demands a narrower inhomogeneous line in order to observe the broad  resonance. Although beyond the scope of this work, the eigenvalues for $\nu \gtrapprox 1$ can be computed numerically using Eqs.  \eqref{eq:consistency_condtion} and \eqref{eq:general-eigenvalue} (see discussion in Appendix \ref{app:eigval_approx}) and substituted in $\eqref{eq:transmission}$ to obtain transmission through a spatially extended and inhomogeneously broadened sample.}  Presently, for a single spatially localized ensemble satisfying \eqref{eq:position_inhomo_limit} and \eqref{eq:freq_inhomo_limit}, one may form the single collective spin (blue emitter, Fig.~\ref{fig:schematic}) operator $\hat{\mathfrak{B}}^{-} = \frac{1}{\sqrt{m}}\sum_{p=1}^{m}\hat{\mathcal{B}}_{p}^{-}$, to be treated as a single optical element. As detailed in Appendix \ref{app:dark_coupling}, a total approximate rate of loss of the coherence $\hat{\mathfrak{B}}^{-}$ via coupling to narrow resonances within the ensemble is given by $\eta(N) = \frac{N\Gamma_{\text{1D}}}{2}\left(\frac{\delta z}{\lambda}\right)$. For $\delta z$ considered here, this rate is at least an order of magnitude smaller than that of the dominant collective decay and interactions through the waveguide at rate $O( N\Gamma_{\text{1D}}/2)$. 
\section{Emulation of CQED}
\begin{figure*}[htb!]
    \centering
\includegraphics[width=\linewidth]{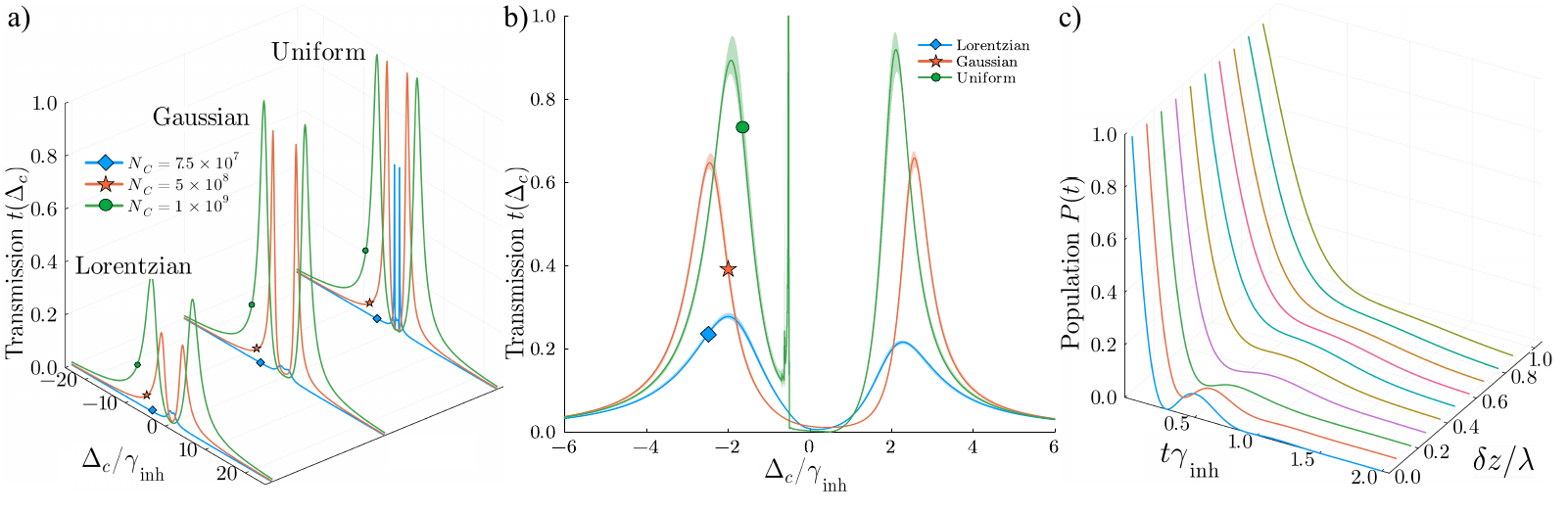}
    \caption{Strong coupling in the emitter-based cavity system. (a) Development of the peak splitting for side-illumination with increasing emitter number $N_{C} = 2N_{Q}$ in the cavity ensembles. We take $\gamma_{\text{inh}}/(2\pi) = 10$GHz, $\Gamma_{\text{1D}}/(2\pi) = \Gamma'/(2\pi) = 100$Hz, and $\delta z = 0$ within each ensemble. (b) Transmission spectrum for the side-illumination scheme~\cite{mirhosseini2019cavity}.
    {We consider $m=10^3$ positional bins for each spatially localized ensemble and show for one realization of positions {uniformly distributed} over a width $\delta z = 0.1\lambda$ within each ensemble.} Parameters are $\gamma_{\text{inh}}/(2\pi) = 10$GHz, $2N_{Q} = N_{C} = 4\times 10^8.$ The shaded regions give the bounds of transmission obtained over 100 realizations of positions.
    The qubit ensemble is additionally detuned (in practice using, e.g., surface acoustic waves~\cite{ohta2021rare,ohta2024acoustic}) to  counter the mirror-qubit detuning that arises according to according to Eq. \ref{eq:perturb_lines}. (c) Rabi oscillations of the qubit population {$P = |\mathfrak{B}_{Q}^{-}|^2$} with $m=10^3$ positional bins sampled from uniform distributions of width $\delta z$.
    A single realization of positions is chosen in each case. Parameters are identical to (b), except $2N_{Q} = N_{C} = 1\times 10^9.$}
    \label{fig:strong_coupling}
\end{figure*}
In this case, coherent interactions between spatially distinct and localized ensembles are possible.
The combined use of localized emitter ensembles satisfying the collective threshold condition \eqref{eq:freq_inhomo_limit} and the single-resonance condition \eqref{eq:position_inhomo_limit} can emulate two mirrors and a qubit placed along the waveguide, forming an \textit{in situ} optical cavity~\cite{mirhosseini2019cavity}.{However, the narrow polariton lines are in general sensitive to small positional fluctuations. We here extend this analysis to both qubit and mirror ensembles featuring inhomogeneous broadening and finite spatial extent.}  
Assume $N_{\text{Q}}$ emitters in a designated qubit ensemble with coherence {$\mathfrak{B}_{Q}^{-}$, and $N_{C}$} emitters each in two designated cavity mirror ensembles {with coherences $\mathfrak{B}_{Cl}^{-}$ 
$(l=1,2)$, with $\delta z = 0$ within each for the moment}. When the mirror ensembles are spaced $\lambda/2 + r\lambda$ (integer $r$) with the qubit ensemble at their midpoint, the qubit interaction with the photonic component of the eigenstates formed by the mirrors is Hamiltonian.  Assuming Lorentzian broadening for illustration and forming the cavity coherence  ${\mathfrak{B}}_{C}^{-} = \frac{1}{\sqrt{2}}\left({\mathfrak{B}}_{C1}^{-} - {\mathfrak{B}}_{C2}^{-}\right)$, the equations of motion describing CQED are obtained as initially proposed in~\cite{chang2012cavity}:
\begin{eqnarray}
    \dot{\mathfrak{B}}_{Q}^{-} &= [\mathrm{i}\Delta_{c} - (\frac{N_{Q}\Gamma_{\text{1D}} + \gamma_{\text{inh}}}{2})]\mathfrak{B}_{Q}^{-}  + \mathrm{i}\sqrt{2N_{Q}N_{C}}\frac{\Gamma_{\text{1D}}}{2}\mathfrak{B}_{C}^{-}, \nonumber \\
    \dot{\mathfrak{B}}_{C}^{-} & \hspace{-40pt} 
 =  (\mathrm{i}\Delta_{c} -  \frac{\gamma_{\text{inh}}}{2})\mathfrak{B}_{C}^{-} + \mathrm{i}\sqrt{2N_{Q}N_{C}}\frac{\Gamma_{\text{1D}}}{2}\mathfrak{B}_{Q}^{-}.
 \label{eq:cavity-qed}
\end{eqnarray}
The analogy to CQED is made with qubit decay rate $\gamma$, cavity decay rate $\kappa$, and coupling $g$ respectively:
\begin{equation}	  \label{eq:cavity_qed:params}
	\gamma = N_{Q}\Gamma_{\text{1D}} + \gamma_{\text{inh}} \ \ \ \ \ \kappa =  \gamma_{\text{inh}}, \ \ \ \ \ g = \sqrt{\frac{N_{Q}N_{C}}{2}}\Gamma_{\text{1D}} .\nonumber
\end{equation}
Notably, for $N_{q} \sim N_{c} \sim N$ we have additionally have $g\sim N\Gamma_{\text{1D}}$, a square root enhancement in $N$ over conventional cavities~\cite{diniz2011strongly}. In the strong-coupling regime, the two eigenvalues of the linear system \eqref{eq:cavity-qed},
\begin{eqnarray}
    \Lambda^{\pm} = \pm \frac{\Gamma_{\text{1D}}}{2}\sqrt{8N_{Q}N_{C} - N_{Q}^2} - \mathrm{i}\left(\frac{N_{Q}\Gamma_{\text{1D}}}{4} + \frac{\gamma_{\text{inh}}}{2}\right),
\end{eqnarray}
are found. {For finite-variance mirror-symmetric inhomogeneous lines the cavity protection effect $\gamma_{\text{inh}} \to \pi \mathfrak{R}[\Lambda^{\pm}]^2\rho\left(\mathfrak{R}[\Lambda^{\pm}]\right) + \Gamma'/2,$ is observed,  limiting $\gamma$ and $\kappa$ to $N_{Q}\Gamma_{\text{1D}} + \Gamma'$ and $\Gamma'$ respectively}.
Whilst for $\Gamma' \ll \gamma_{\text{inh}}, N\Gamma_{\text{1D}}$ the generic conditions to be well within the strong coupling regime approximately read $\sqrt{N_{C}N_{Q}} \gg \frac{\gamma_{\text{inh}}}{\Gamma_{\text{1D}}}, 2N_{C} \geq N_{Q},$ there is significant variation in the onset of peak visibility for differing spectral distributions. Applying a side illumination scheme to avoid exciting the broad mirror resonance~\cite{mirhosseini2019cavity}, this effect can be observed in Fig.~\ref{fig:strong_coupling}(a), first considering the case of $\delta z = 0$ in each ensemble.
For parameters $\gamma_{\text{inh}}/(2\pi) = 10$GHz, and $\Gamma_{\text{1D}}/(2\pi) = \Gamma'/(2\pi) = 100$Hz corresponding to, say, the optical Erbium transition $Y_{1} \to Z_{1}$ implanted into YSO~\cite{ohta2021rare} or grown in rare-earth oxides~\cite{yasui2022creation,pak2022long}, peaks are established in the Gaussian and uniform case for individual ensemble numbers as low as $N = 10^8$, which for emitters localized in a region of size $\delta z \times \lambda^2 = 0.1\lambda \times \lambda^2 = 0.1\lambda^3$ corresponds to doping concentrations below the achievable $10^{22}$cm$^{-3}$~\cite{xu2021low}. {When reintroducing positional inhomogeneity within the qubit and mirror ensembles, one must be careful to choose $N_{Q}, N_{c}$ so that $\eta(2N_{c}),\eta(N_{q}) \lessapprox g,$ which bounds $N_{c},N_{q}$ from below. That is, the coupling between symmetric qubit and mirror ensemble excitations should exceed the ensemble decoherence of qubit and mirrors due to finite spatial extent.} Choosing $2N_{q} = N_{c}$ to satisfy this condition, we see in Fig.~\ref{fig:strong_coupling}(b) for $\delta z = 0.1\lambda$ that high visibility peaks are still retained and the effect of finite spatial extent is limited to narrow central resonances. The corresponding Rabi oscillations of the qubit ensemble in Fig.~\ref{fig:strong_coupling}(c) are preserved for regimes of smaller spatial extent even in the lowest-fidelity Lorentzian case and illustrate coherent population transfer between the qubit and cavity modes. Note that by using, e.g., photonic crystal waveguides~\cite{viasnoff2005spontaneous,rao2007single} or plasmonics~\cite{Gsken2023} to enhance $\Gamma_{\text{1D}},$ the density requirements for constant peak visibility can be reduced by 1-2 orders of magnitude, or peak visibility enhanced for constant $N$. With optical wavelengths on the order of $\mu$m and possible waveguide lengths on the order of mm, the local density within a single ensemble can be reduced even further orders of magnitude by employing commensurate ensembles along the waveguide as a single unit~\cite{chang2012cavity}. For microwave transitions the long wavelength allows orders of magnitude more emitters in a given wavelength, and also offers a promising platform for observing strong coupling exclusively amongst solid-state emitters.


\section{Conclusion}
In this work we quantitatively investigated when collective spins formed from spatially localized and inhomogeneously broadened rare-earth ions can be employed as coherent and spectrally tailorable emitters in the paradigm of waveguide QED and in currently accessible and near-term experimental platforms. When the linewidth of a collective resonance -- dependent on ensemble spatial extent -- exceeds that of inhomogeneous broadening, the single broad collective resonance becomes accessible in the spectrum for spatial extents observed in existing platforms based on rare-earth ions, allowing for the formation of coherent optical elements. As such, when localized ensembles are combined, the strong coupling regime of CQED can be readily accessed despite positional disorder, and with relatively low emitter concentrations if spectral hole burning is additionally employed. These results suggest the potential of the inhomogeneous ensembles for coherent interactions in optical waveguides beyond extended bulk applications, and advance the theory of mesoscopic systems of optical emitters.


\begin{acknowledgments}
The authors are grateful for valuable discussions with K. Azuma. Simulations in this work were performed with the aid of the open-source \verb|Julia| package DifferentialEquations.jl~\cite{rackauckas2017differentialequations}. This work was supported by JSPS KAKENHI, Grants No. JP23H01112. 
\end{acknowledgments}


\providecommand{\noopsort}[1]{}\providecommand{\singleletter}[1]{#1}%
%


\begin{thebibliography}{55}%
\makeatletter
\providecommand \@ifxundefined [1]{%
 \@ifx{#1\undefined}
}%
\providecommand \@ifnum [1]{%
 \ifnum #1\expandafter \@firstoftwo
 \else \expandafter \@secondoftwo
 \fi
}%
\providecommand \@ifx [1]{%
 \ifx #1\expandafter \@firstoftwo
 \else \expandafter \@secondoftwo
 \fi
}%
\providecommand \natexlab [1]{#1}%
\providecommand \enquote  [1]{``#1''}%
\providecommand \bibnamefont  [1]{#1}%
\providecommand \bibfnamefont [1]{#1}%
\providecommand \citenamefont [1]{#1}%
\providecommand \href@noop [0]{\@secondoftwo}%
\providecommand \href [0]{\begingroup \@sanitize@url \@href}%
\providecommand \@href[1]{\@@startlink{#1}\@@href}%
\providecommand \@@href[1]{\endgroup#1\@@endlink}%
\providecommand \@sanitize@url [0]{\catcode `\\12\catcode `\$12\catcode
  `\&12\catcode `\#12\catcode `\^12\catcode `\_12\catcode `\%12\relax}%
\providecommand \@@startlink[1]{}%
\providecommand \@@endlink[0]{}%
\providecommand \url  [0]{\begingroup\@sanitize@url \@url }%
\providecommand \@url [1]{\endgroup\@href {#1}{\urlprefix }}%
\providecommand \urlprefix  [0]{URL }%
\providecommand \Eprint [0]{\href }%
\providecommand \doibase [0]{https://doi.org/}%
\providecommand \selectlanguage [0]{\@gobble}%
\providecommand \bibinfo  [0]{\@secondoftwo}%
\providecommand \bibfield  [0]{\@secondoftwo}%
\providecommand \translation [1]{[#1]}%
\providecommand \BibitemOpen [0]{}%
\providecommand \bibitemStop [0]{}%
\providecommand \bibitemNoStop [0]{.\EOS\space}%
\providecommand \EOS [0]{\spacefactor3000\relax}%
\providecommand \BibitemShut  [1]{\csname bibitem#1\endcsname}%
\let\auto@bib@innerbib\@empty
\bibitem [{\citenamefont {Kurizki}\ \emph {et~al.}(2015)\citenamefont
  {Kurizki}, \citenamefont {Bertet}, \citenamefont {Kubo}, \citenamefont
  {M{\o}lmer}, \citenamefont {Petrosyan}, \citenamefont {Rabl},\ and\
  \citenamefont {Schmiedmayer}}]{Kurizki_2015}%
  \BibitemOpen
  \bibfield  {author} {\bibinfo {author} {\bibfnamefont {G.}~\bibnamefont
  {Kurizki}}, \bibinfo {author} {\bibfnamefont {P.}~\bibnamefont {Bertet}},
  \bibinfo {author} {\bibfnamefont {Y.}~\bibnamefont {Kubo}}, \bibinfo {author}
  {\bibfnamefont {K.}~\bibnamefont {M{\o}lmer}}, \bibinfo {author}
  {\bibfnamefont {D.}~\bibnamefont {Petrosyan}}, \bibinfo {author}
  {\bibfnamefont {P.}~\bibnamefont {Rabl}},\ and\ \bibinfo {author}
  {\bibfnamefont {J.}~\bibnamefont {Schmiedmayer}},\ }\bibfield  {title}
  {\bibinfo {title} {Quantum technologies with hybrid systems},\ }\href
  {https://doi.org/10.1073/pnas.1419326112} {\bibfield  {journal} {\bibinfo
  {journal} {Proc. Nat. Acad. Sci. USA}\ }\textbf {\bibinfo {volume} {112}},\
  \bibinfo {pages} {3866} (\bibinfo {year} {2015})}\BibitemShut {NoStop}%
\bibitem [{\citenamefont {Serrano}\ \emph {et~al.}(2022)\citenamefont
  {Serrano}, \citenamefont {Kuppusamy}, \citenamefont {Heinrich}, \citenamefont
  {Fuhr}, \citenamefont {Hunger}, \citenamefont {Ruben},\ and\ \citenamefont
  {Goldner}}]{serrano2022ultra}%
  \BibitemOpen
  \bibfield  {author} {\bibinfo {author} {\bibfnamefont {D.}~\bibnamefont
  {Serrano}}, \bibinfo {author} {\bibfnamefont {S.~K.}\ \bibnamefont
  {Kuppusamy}}, \bibinfo {author} {\bibfnamefont {B.}~\bibnamefont {Heinrich}},
  \bibinfo {author} {\bibfnamefont {O.}~\bibnamefont {Fuhr}}, \bibinfo {author}
  {\bibfnamefont {D.}~\bibnamefont {Hunger}}, \bibinfo {author} {\bibfnamefont
  {M.}~\bibnamefont {Ruben}},\ and\ \bibinfo {author} {\bibfnamefont
  {P.}~\bibnamefont {Goldner}},\ }\bibfield  {title} {\bibinfo {title}
  {Ultra-narrow optical linewidths in rare-earth molecular crystals},\ }\href
  {https://doi.org/10.1038/s41586-021-04316-2} {\bibfield  {journal} {\bibinfo
  {journal} {Nature}\ }\textbf {\bibinfo {volume} {603}},\ \bibinfo {pages}
  {241} (\bibinfo {year} {2022})}\BibitemShut {NoStop}%
\bibitem [{\citenamefont {Ohta}\ \emph {et~al.}(2021)\citenamefont {Ohta},
  \citenamefont {Herpin}, \citenamefont {Bastidas}, \citenamefont {Tawara},
  \citenamefont {Yamaguchi},\ and\ \citenamefont {Okamoto}}]{ohta2021rare}%
  \BibitemOpen
  \bibfield  {author} {\bibinfo {author} {\bibfnamefont {R.}~\bibnamefont
  {Ohta}}, \bibinfo {author} {\bibfnamefont {L.}~\bibnamefont {Herpin}},
  \bibinfo {author} {\bibfnamefont {V.~M.}\ \bibnamefont {Bastidas}}, \bibinfo
  {author} {\bibfnamefont {T.}~\bibnamefont {Tawara}}, \bibinfo {author}
  {\bibfnamefont {H.}~\bibnamefont {Yamaguchi}},\ and\ \bibinfo {author}
  {\bibfnamefont {H.}~\bibnamefont {Okamoto}},\ }\bibfield  {title} {\bibinfo
  {title} {Rare-earth-mediated optomechanical system in the reversed
  dissipation regime},\ }\href {https://doi.org/10.1103/physrevlett.126.047404}
  {\bibfield  {journal} {\bibinfo  {journal} {Phys. Rev. Lett.}\ }\textbf
  {\bibinfo {volume} {126}} (\bibinfo {year} {2021})}\BibitemShut {NoStop}%
\bibitem [{\citenamefont {Yasui}\ \emph {et~al.}(2022)\citenamefont {Yasui},
  \citenamefont {Hiraishi}, \citenamefont {Ishizawa}, \citenamefont {Omi},
  \citenamefont {Inaba}, \citenamefont {Xu}, \citenamefont {Kaji},
  \citenamefont {Adachi},\ and\ \citenamefont {Tawara}}]{yasui2022creation}%
  \BibitemOpen
  \bibfield  {author} {\bibinfo {author} {\bibfnamefont {S.}~\bibnamefont
  {Yasui}}, \bibinfo {author} {\bibfnamefont {M.}~\bibnamefont {Hiraishi}},
  \bibinfo {author} {\bibfnamefont {A.}~\bibnamefont {Ishizawa}}, \bibinfo
  {author} {\bibfnamefont {H.}~\bibnamefont {Omi}}, \bibinfo {author}
  {\bibfnamefont {T.}~\bibnamefont {Inaba}}, \bibinfo {author} {\bibfnamefont
  {X.}~\bibnamefont {Xu}}, \bibinfo {author} {\bibfnamefont {R.}~\bibnamefont
  {Kaji}}, \bibinfo {author} {\bibfnamefont {S.}~\bibnamefont {Adachi}},\ and\
  \bibinfo {author} {\bibfnamefont {T.}~\bibnamefont {Tawara}},\ }\bibfield
  {title} {\bibinfo {title} {Creation of a high-resolution atomic frequency
  comb and optimization of the pulse sequence for high-efficiency quantum
  memory in $^{167}$ {E}r:{Y}$_{2}${S}i{O}$_{5}$},\ }\href
  {https://doi.org/10.1364/optcon.457429} {\bibfield  {journal} {\bibinfo
  {journal} {Opt. Cont.}\ }\textbf {\bibinfo {volume} {1}},\ \bibinfo {pages}
  {1896} (\bibinfo {year} {2022})}\BibitemShut {NoStop}%
\bibitem [{\citenamefont {Lago-Rivera}\ \emph {et~al.}(2021)\citenamefont
  {Lago-Rivera}, \citenamefont {Grandi}, \citenamefont {Rakonjac},
  \citenamefont {Seri},\ and\ \citenamefont {de~Riedmatten}}]{lago2021telecom}%
  \BibitemOpen
  \bibfield  {author} {\bibinfo {author} {\bibfnamefont {D.}~\bibnamefont
  {Lago-Rivera}}, \bibinfo {author} {\bibfnamefont {S.}~\bibnamefont {Grandi}},
  \bibinfo {author} {\bibfnamefont {J.~V.}\ \bibnamefont {Rakonjac}}, \bibinfo
  {author} {\bibfnamefont {A.}~\bibnamefont {Seri}},\ and\ \bibinfo {author}
  {\bibfnamefont {H.}~\bibnamefont {de~Riedmatten}},\ }\bibfield  {title}
  {\bibinfo {title} {Telecom-heralded entanglement between multimode
  solid-state quantum memories},\ }\href
  {https://doi.org/10.1038/s41586-021-03481-8} {\bibfield  {journal} {\bibinfo
  {journal} {Nature}\ }\textbf {\bibinfo {volume} {594}},\ \bibinfo {pages}
  {37} (\bibinfo {year} {2021})}\BibitemShut {NoStop}%
\bibitem [{\citenamefont {Zhong}\ \emph {et~al.}(2017)\citenamefont {Zhong},
  \citenamefont {Kindem}, \citenamefont {Rochman},\ and\ \citenamefont
  {Faraon}}]{zhong2017interfacing}%
  \BibitemOpen
  \bibfield  {author} {\bibinfo {author} {\bibfnamefont {T.}~\bibnamefont
  {Zhong}}, \bibinfo {author} {\bibfnamefont {J.~M.}\ \bibnamefont {Kindem}},
  \bibinfo {author} {\bibfnamefont {J.}~\bibnamefont {Rochman}},\ and\ \bibinfo
  {author} {\bibfnamefont {A.}~\bibnamefont {Faraon}},\ }\bibfield  {title}
  {\bibinfo {title} {Interfacing broadband photonic qubits to on-chip
  cavity-protected rare-earth ensembles},\ }\href
  {https://doi.org/10.1038/ncomms14107} {\bibfield  {journal} {\bibinfo
  {journal} {Nat. Comm.}\ }\textbf {\bibinfo {volume} {8}} (\bibinfo {year}
  {2017})}\BibitemShut {NoStop}%
\bibitem [{\citenamefont {Zhu}\ \emph {et~al.}(2011)\citenamefont {Zhu},
  \citenamefont {Saito}, \citenamefont {Kemp}, \citenamefont {Kakuyanagi},
  \citenamefont {ichi Karimoto}, \citenamefont {Nakano}, \citenamefont {Munro},
  \citenamefont {Tokura}, \citenamefont {Everitt}, \citenamefont {Nemoto},
  \citenamefont {Kasu}, \citenamefont {Mizuochi},\ and\ \citenamefont
  {Semba}}]{zhu2011coherent}%
  \BibitemOpen
  \bibfield  {author} {\bibinfo {author} {\bibfnamefont {X.}~\bibnamefont
  {Zhu}}, \bibinfo {author} {\bibfnamefont {S.}~\bibnamefont {Saito}}, \bibinfo
  {author} {\bibfnamefont {A.}~\bibnamefont {Kemp}}, \bibinfo {author}
  {\bibfnamefont {K.}~\bibnamefont {Kakuyanagi}}, \bibinfo {author}
  {\bibfnamefont {S.}~\bibnamefont {ichi Karimoto}}, \bibinfo {author}
  {\bibfnamefont {H.}~\bibnamefont {Nakano}}, \bibinfo {author} {\bibfnamefont
  {W.~J.}\ \bibnamefont {Munro}}, \bibinfo {author} {\bibfnamefont
  {Y.}~\bibnamefont {Tokura}}, \bibinfo {author} {\bibfnamefont {M.~S.}\
  \bibnamefont {Everitt}}, \bibinfo {author} {\bibfnamefont {K.}~\bibnamefont
  {Nemoto}}, \bibinfo {author} {\bibfnamefont {M.}~\bibnamefont {Kasu}},
  \bibinfo {author} {\bibfnamefont {N.}~\bibnamefont {Mizuochi}},\ and\
  \bibinfo {author} {\bibfnamefont {K.}~\bibnamefont {Semba}},\ }\bibfield
  {title} {\bibinfo {title} {Coherent coupling of a superconducting flux qubit
  to an electron spin ensemble in diamond},\ }\href
  {https://doi.org/10.1038/nature10462} {\bibfield  {journal} {\bibinfo
  {journal} {Nature}\ }\textbf {\bibinfo {volume} {478}},\ \bibinfo {pages}
  {221} (\bibinfo {year} {2011})}\BibitemShut {NoStop}%
\bibitem [{\citenamefont {Ams\"{u}ss}\ \emph {et~al.}(2011)\citenamefont
  {Ams\"{u}ss}, \citenamefont {Koller}, \citenamefont {N\"{o}bauer},
  \citenamefont {Putz}, \citenamefont {Rotter}, \citenamefont {Sandner},
  \citenamefont {Schneider}, \citenamefont {Schramb\"{o}ck}, \citenamefont
  {Steinhauser}, \citenamefont {Ritsch}, \citenamefont {Schmiedmayer},\ and\
  \citenamefont {Majer}}]{Amsss2011}%
  \BibitemOpen
  \bibfield  {author} {\bibinfo {author} {\bibfnamefont {R.}~\bibnamefont
  {Ams\"{u}ss}}, \bibinfo {author} {\bibfnamefont {C.}~\bibnamefont {Koller}},
  \bibinfo {author} {\bibfnamefont {T.}~\bibnamefont {N\"{o}bauer}}, \bibinfo
  {author} {\bibfnamefont {S.}~\bibnamefont {Putz}}, \bibinfo {author}
  {\bibfnamefont {S.}~\bibnamefont {Rotter}}, \bibinfo {author} {\bibfnamefont
  {K.}~\bibnamefont {Sandner}}, \bibinfo {author} {\bibfnamefont
  {S.}~\bibnamefont {Schneider}}, \bibinfo {author} {\bibfnamefont
  {M.}~\bibnamefont {Schramb\"{o}ck}}, \bibinfo {author} {\bibfnamefont
  {G.}~\bibnamefont {Steinhauser}}, \bibinfo {author} {\bibfnamefont
  {H.}~\bibnamefont {Ritsch}}, \bibinfo {author} {\bibfnamefont
  {J.}~\bibnamefont {Schmiedmayer}},\ and\ \bibinfo {author} {\bibfnamefont
  {J.}~\bibnamefont {Majer}},\ }\bibfield  {title} {\bibinfo {title} {Cavity
  {QED} with magnetically coupled collective spin states},\ }\href
  {https://doi.org/10.1103/physrevlett.107.060502} {\bibfield  {journal}
  {\bibinfo  {journal} {Phys. Rev. Lett.}\ }\textbf {\bibinfo {volume} {107}}
  (\bibinfo {year} {2011})}\BibitemShut {NoStop}%
\bibitem [{\citenamefont {Lei}\ \emph {et~al.}(2023)\citenamefont {Lei},
  \citenamefont {Fukumori}, \citenamefont {Rochman}, \citenamefont {Zhu},
  \citenamefont {Endres}, \citenamefont {Choi},\ and\ \citenamefont
  {Faraon}}]{Lei2023}%
  \BibitemOpen
  \bibfield  {author} {\bibinfo {author} {\bibfnamefont {M.}~\bibnamefont
  {Lei}}, \bibinfo {author} {\bibfnamefont {R.}~\bibnamefont {Fukumori}},
  \bibinfo {author} {\bibfnamefont {J.}~\bibnamefont {Rochman}}, \bibinfo
  {author} {\bibfnamefont {B.}~\bibnamefont {Zhu}}, \bibinfo {author}
  {\bibfnamefont {M.}~\bibnamefont {Endres}}, \bibinfo {author} {\bibfnamefont
  {J.}~\bibnamefont {Choi}},\ and\ \bibinfo {author} {\bibfnamefont
  {A.}~\bibnamefont {Faraon}},\ }\bibfield  {title} {\bibinfo {title}
  {Many-body cavity quantum electrodynamics with driven inhomogeneous
  emitters},\ }\href {https://doi.org/10.1038/s41586-023-05884-1} {\bibfield
  {journal} {\bibinfo  {journal} {Nature}\ }\textbf {\bibinfo {volume} {617}},\
  \bibinfo {pages} {271} (\bibinfo {year} {2023})}\BibitemShut {NoStop}%
\bibitem [{\citenamefont {Becker}\ \emph {et~al.}(1999)\citenamefont {Becker},
  \citenamefont {Olsson},\ and\ \citenamefont {Simpson}}]{becker1999erbium}%
  \BibitemOpen
  \bibfield  {author} {\bibinfo {author} {\bibfnamefont {P.~M.}\ \bibnamefont
  {Becker}}, \bibinfo {author} {\bibfnamefont {A.~A.}\ \bibnamefont {Olsson}},\
  and\ \bibinfo {author} {\bibfnamefont {J.~R.}\ \bibnamefont {Simpson}},\
  }\href {https://doi.org/10.1016/b978-0-12-084590-3.x5000-5} {\emph {\bibinfo
  {title} {Erbium-doped fiber amplifiers: fundamentals and technology}}}\
  (\bibinfo  {publisher} {Elsevier},\ \bibinfo {year} {1999})\BibitemShut
  {NoStop}%
\bibitem [{\citenamefont {Rinner}\ \emph {et~al.}(2023)\citenamefont {Rinner},
  \citenamefont {Burger}, \citenamefont {Gritsch}, \citenamefont {Schmitt},\
  and\ \citenamefont {Reiserer}}]{rinner2023erbium}%
  \BibitemOpen
  \bibfield  {author} {\bibinfo {author} {\bibfnamefont {S.}~\bibnamefont
  {Rinner}}, \bibinfo {author} {\bibfnamefont {F.}~\bibnamefont {Burger}},
  \bibinfo {author} {\bibfnamefont {A.}~\bibnamefont {Gritsch}}, \bibinfo
  {author} {\bibfnamefont {J.}~\bibnamefont {Schmitt}},\ and\ \bibinfo {author}
  {\bibfnamefont {A.}~\bibnamefont {Reiserer}},\ }\bibfield  {title} {\bibinfo
  {title} {Erbium emitters in commercially fabricated nanophotonic silicon
  waveguides},\ }\href {https://doi.org/10.1515/nanoph-2023-0287} {\bibfield
  {journal} {\bibinfo  {journal} {Nanophotonics}\ } (\bibinfo {year}
  {2023})}\BibitemShut {NoStop}%
\bibitem [{\citenamefont {Weiss}\ \emph {et~al.}(2021)\citenamefont {Weiss},
  \citenamefont {Gritsch}, \citenamefont {Merkel},\ and\ \citenamefont
  {Reiserer}}]{Weiss2021}%
  \BibitemOpen
  \bibfield  {author} {\bibinfo {author} {\bibfnamefont {L.}~\bibnamefont
  {Weiss}}, \bibinfo {author} {\bibfnamefont {A.}~\bibnamefont {Gritsch}},
  \bibinfo {author} {\bibfnamefont {B.}~\bibnamefont {Merkel}},\ and\ \bibinfo
  {author} {\bibfnamefont {A.}~\bibnamefont {Reiserer}},\ }\bibfield  {title}
  {\bibinfo {title} {Erbium dopants in nanophotonic silicon waveguides},\
  }\href {https://doi.org/10.1364/optica.413330} {\bibfield  {journal}
  {\bibinfo  {journal} {Optica}\ }\textbf {\bibinfo {volume} {8}},\ \bibinfo
  {pages} {40} (\bibinfo {year} {2021})}\BibitemShut {NoStop}%
\bibitem [{\citenamefont {Mor}\ \emph {et~al.}(2022)\citenamefont {Mor},
  \citenamefont {Ohana}, \citenamefont {Borne}, \citenamefont {Diskin-Posner},
  \citenamefont {Asher}, \citenamefont {Yaffe}, \citenamefont {Shanzer},\ and\
  \citenamefont {Dayan}}]{Mor2022}%
  \BibitemOpen
  \bibfield  {author} {\bibinfo {author} {\bibfnamefont {O.~E.}\ \bibnamefont
  {Mor}}, \bibinfo {author} {\bibfnamefont {T.}~\bibnamefont {Ohana}}, \bibinfo
  {author} {\bibfnamefont {A.}~\bibnamefont {Borne}}, \bibinfo {author}
  {\bibfnamefont {Y.}~\bibnamefont {Diskin-Posner}}, \bibinfo {author}
  {\bibfnamefont {M.}~\bibnamefont {Asher}}, \bibinfo {author} {\bibfnamefont
  {O.}~\bibnamefont {Yaffe}}, \bibinfo {author} {\bibfnamefont
  {A.}~\bibnamefont {Shanzer}},\ and\ \bibinfo {author} {\bibfnamefont
  {B.}~\bibnamefont {Dayan}},\ }\bibfield  {title} {\bibinfo {title} {Tapered
  optical fibers coated with rare-earth complexes for quantum applications},\
  }\href {https://doi.org/10.1021/acsphotonics.2c00330} {\bibfield  {journal}
  {\bibinfo  {journal} {{ACS} Phot.}\ }\textbf {\bibinfo {volume} {9}},\
  \bibinfo {pages} {2676} (\bibinfo {year} {2022})}\BibitemShut {NoStop}%
\bibitem [{\citenamefont {Sipahigil}\ \emph {et~al.}(2016)\citenamefont
  {Sipahigil}, \citenamefont {Evans}, \citenamefont {Sukachev}, \citenamefont
  {Burek}, \citenamefont {Borregaard}, \citenamefont {Bhaskar}, \citenamefont
  {Nguyen}, \citenamefont {Pacheco}, \citenamefont {Atikian}, \citenamefont
  {Meuwly}, \citenamefont {Camacho}, \citenamefont {Jelezko}, \citenamefont
  {Bielejec}, \citenamefont {Park}, \citenamefont {Lon{\v{c}}ar},\ and\
  \citenamefont {Lukin}}]{Sipahigil2016}%
  \BibitemOpen
  \bibfield  {author} {\bibinfo {author} {\bibfnamefont {A.}~\bibnamefont
  {Sipahigil}}, \bibinfo {author} {\bibfnamefont {R.~E.}\ \bibnamefont
  {Evans}}, \bibinfo {author} {\bibfnamefont {D.~D.}\ \bibnamefont {Sukachev}},
  \bibinfo {author} {\bibfnamefont {M.~J.}\ \bibnamefont {Burek}}, \bibinfo
  {author} {\bibfnamefont {J.}~\bibnamefont {Borregaard}}, \bibinfo {author}
  {\bibfnamefont {M.~K.}\ \bibnamefont {Bhaskar}}, \bibinfo {author}
  {\bibfnamefont {C.~T.}\ \bibnamefont {Nguyen}}, \bibinfo {author}
  {\bibfnamefont {J.~L.}\ \bibnamefont {Pacheco}}, \bibinfo {author}
  {\bibfnamefont {H.~A.}\ \bibnamefont {Atikian}}, \bibinfo {author}
  {\bibfnamefont {C.}~\bibnamefont {Meuwly}}, \bibinfo {author} {\bibfnamefont
  {R.~M.}\ \bibnamefont {Camacho}}, \bibinfo {author} {\bibfnamefont
  {F.}~\bibnamefont {Jelezko}}, \bibinfo {author} {\bibfnamefont
  {E.}~\bibnamefont {Bielejec}}, \bibinfo {author} {\bibfnamefont
  {H.}~\bibnamefont {Park}}, \bibinfo {author} {\bibfnamefont {M.}~\bibnamefont
  {Lon{\v{c}}ar}},\ and\ \bibinfo {author} {\bibfnamefont {M.~D.}\ \bibnamefont
  {Lukin}},\ }\bibfield  {title} {\bibinfo {title} {An integrated diamond
  nanophotonics platform for quantum-optical networks},\ }\href
  {https://doi.org/10.1126/science.aah6875} {\bibfield  {journal} {\bibinfo
  {journal} {Science}\ }\textbf {\bibinfo {volume} {354}},\ \bibinfo {pages}
  {847} (\bibinfo {year} {2016})}\BibitemShut {NoStop}%
\bibitem [{\citenamefont {Faraon}\ \emph {et~al.}(2011)\citenamefont {Faraon},
  \citenamefont {Majumdar}, \citenamefont {Englund}, \citenamefont {Kim},
  \citenamefont {Bajcsy},\ and\ \citenamefont
  {Vu{\v{c}}kovi{\'{c}}}}]{Faraon2011}%
  \BibitemOpen
  \bibfield  {author} {\bibinfo {author} {\bibfnamefont {A.}~\bibnamefont
  {Faraon}}, \bibinfo {author} {\bibfnamefont {A.}~\bibnamefont {Majumdar}},
  \bibinfo {author} {\bibfnamefont {D.}~\bibnamefont {Englund}}, \bibinfo
  {author} {\bibfnamefont {E.}~\bibnamefont {Kim}}, \bibinfo {author}
  {\bibfnamefont {M.}~\bibnamefont {Bajcsy}},\ and\ \bibinfo {author}
  {\bibfnamefont {J.}~\bibnamefont {Vu{\v{c}}kovi{\'{c}}}},\ }\bibfield
  {title} {\bibinfo {title} {Integrated quantum optical networks based on
  quantum dots and photonic crystals},\ }\href
  {https://doi.org/10.1088/1367-2630/13/5/055025} {\bibfield  {journal}
  {\bibinfo  {journal} {New J. Phys.}\ }\textbf {\bibinfo {volume} {13}},\
  \bibinfo {pages} {055025} (\bibinfo {year} {2011})}\BibitemShut {NoStop}%
\bibitem [{\citenamefont {Choi}\ \emph {et~al.}(2008)\citenamefont {Choi},
  \citenamefont {Deng}, \citenamefont {Laurat},\ and\ \citenamefont
  {Kimble}}]{Choi2008}%
  \BibitemOpen
  \bibfield  {author} {\bibinfo {author} {\bibfnamefont {K.~S.}\ \bibnamefont
  {Choi}}, \bibinfo {author} {\bibfnamefont {H.}~\bibnamefont {Deng}}, \bibinfo
  {author} {\bibfnamefont {J.}~\bibnamefont {Laurat}},\ and\ \bibinfo {author}
  {\bibfnamefont {H.~J.}\ \bibnamefont {Kimble}},\ }\bibfield  {title}
  {\bibinfo {title} {Mapping photonic entanglement into and out of a quantum
  memory},\ }\href {https://doi.org/10.1038/nature06670} {\bibfield  {journal}
  {\bibinfo  {journal} {Nature}\ }\textbf {\bibinfo {volume} {452}},\ \bibinfo
  {pages} {67} (\bibinfo {year} {2008})}\BibitemShut {NoStop}%
\bibitem [{\citenamefont {Julsgaard}\ \emph {et~al.}(2013)\citenamefont
  {Julsgaard}, \citenamefont {Grezes}, \citenamefont {Bertet},\ and\
  \citenamefont {M{\o}lmer}}]{Julsgaard2013}%
  \BibitemOpen
  \bibfield  {author} {\bibinfo {author} {\bibfnamefont {B.}~\bibnamefont
  {Julsgaard}}, \bibinfo {author} {\bibfnamefont {C.}~\bibnamefont {Grezes}},
  \bibinfo {author} {\bibfnamefont {P.}~\bibnamefont {Bertet}},\ and\ \bibinfo
  {author} {\bibfnamefont {K.}~\bibnamefont {M{\o}lmer}},\ }\bibfield  {title}
  {\bibinfo {title} {Quantum memory for microwave photons in an inhomogeneously
  broadened spin ensemble},\ }\href
  {https://doi.org/10.1103/physrevlett.110.250503} {\bibfield  {journal}
  {\bibinfo  {journal} {Phys. Rev. Lett.}\ }\textbf {\bibinfo {volume} {110}}
  (\bibinfo {year} {2013})}\BibitemShut {NoStop}%
\bibitem [{\citenamefont {Afzelius}\ \emph {et~al.}(2009)\citenamefont
  {Afzelius}, \citenamefont {Simon}, \citenamefont {de~Riedmatten},\ and\
  \citenamefont {Gisin}}]{Afzelius_2009}%
  \BibitemOpen
  \bibfield  {author} {\bibinfo {author} {\bibfnamefont {M.}~\bibnamefont
  {Afzelius}}, \bibinfo {author} {\bibfnamefont {C.}~\bibnamefont {Simon}},
  \bibinfo {author} {\bibfnamefont {H.}~\bibnamefont {de~Riedmatten}},\ and\
  \bibinfo {author} {\bibfnamefont {N.}~\bibnamefont {Gisin}},\ }\bibfield
  {title} {\bibinfo {title} {Multimode quantum memory based on atomic frequency
  combs},\ }\href {https://doi.org/10.1103%2Fphysreva.79.052329} {\bibfield
  {journal} {\bibinfo  {journal} {Phys. Rev. A}\ }\textbf {\bibinfo {volume}
  {79}} (\bibinfo {year} {2009})}\BibitemShut {NoStop}%
\bibitem [{\citenamefont {Sheremet}\ \emph {et~al.}(2023)\citenamefont
  {Sheremet}, \citenamefont {Petrov}, \citenamefont {Iorsh}, \citenamefont
  {Poshakinskiy},\ and\ \citenamefont {Poddubny}}]{sheremet2023waveguide}%
  \BibitemOpen
  \bibfield  {author} {\bibinfo {author} {\bibfnamefont {A.~S.}\ \bibnamefont
  {Sheremet}}, \bibinfo {author} {\bibfnamefont {M.~I.}\ \bibnamefont
  {Petrov}}, \bibinfo {author} {\bibfnamefont {I.~V.}\ \bibnamefont {Iorsh}},
  \bibinfo {author} {\bibfnamefont {A.~V.}\ \bibnamefont {Poshakinskiy}},\ and\
  \bibinfo {author} {\bibfnamefont {A.~N.}\ \bibnamefont {Poddubny}},\
  }\bibfield  {title} {\bibinfo {title} {Waveguide quantum electrodynamics:
  collective radiance and photon-photon correlations},\ }\href
  {https://doi.org/10.1103/revmodphys.95.015002} {\bibfield  {journal}
  {\bibinfo  {journal} {Rev. of Mod. Phys.}\ }\textbf {\bibinfo {volume}
  {95}},\ \bibinfo {pages} {015002} (\bibinfo {year} {2023})}\BibitemShut
  {NoStop}%
\bibitem [{\citenamefont {Kling}\ and\ \citenamefont
  {Hosseini}(2023)}]{kling2023characteristics}%
  \BibitemOpen
  \bibfield  {author} {\bibinfo {author} {\bibfnamefont {T.}~\bibnamefont
  {Kling}}\ and\ \bibinfo {author} {\bibfnamefont {M.}~\bibnamefont
  {Hosseini}},\ }\bibfield  {title} {\bibinfo {title} {Characteristics of 1{D}
  ordered arrays of optical centers in solid-state photonics},\ }\href
  {https://doi.org/10.1088/2515-7647/acccc3} {\bibfield  {journal} {\bibinfo
  {journal} {J. Phys: Phot.}\ }\textbf {\bibinfo {volume} {5}},\ \bibinfo
  {pages} {024003} (\bibinfo {year} {2023})}\BibitemShut {NoStop}%
\bibitem [{\citenamefont {Song}\ \emph {et~al.}(2021)\citenamefont {Song},
  \citenamefont {Guo}, \citenamefont {Nie}, \citenamefont {Kwek},\ and\
  \citenamefont {Long}}]{Song:21}%
  \BibitemOpen
  \bibfield  {author} {\bibinfo {author} {\bibfnamefont {G.-Z.}\ \bibnamefont
  {Song}}, \bibinfo {author} {\bibfnamefont {J.-L.}\ \bibnamefont {Guo}},
  \bibinfo {author} {\bibfnamefont {W.}~\bibnamefont {Nie}}, \bibinfo {author}
  {\bibfnamefont {L.-C.}\ \bibnamefont {Kwek}},\ and\ \bibinfo {author}
  {\bibfnamefont {G.-L.}\ \bibnamefont {Long}},\ }\bibfield  {title} {\bibinfo
  {title} {Optical properties of a waveguide-mediated chain of randomly
  positioned atoms},\ }\href {https://doi.org/10.1364/OE.409471} {\bibfield
  {journal} {\bibinfo  {journal} {Opt. Express}\ }\textbf {\bibinfo {volume}
  {29}},\ \bibinfo {pages} {1903} (\bibinfo {year} {2021})}\BibitemShut
  {NoStop}%
\bibitem [{\citenamefont {Ruostekoski}\ and\ \citenamefont
  {Javanainen}(2016)}]{Ruostekoski2016}%
  \BibitemOpen
  \bibfield  {author} {\bibinfo {author} {\bibfnamefont {J.}~\bibnamefont
  {Ruostekoski}}\ and\ \bibinfo {author} {\bibfnamefont {J.}~\bibnamefont
  {Javanainen}},\ }\bibfield  {title} {\bibinfo {title} {Emergence of
  correlated optics in one-dimensional waveguides for classical and quantum
  atomic gases},\ }\href {https://doi.org/10.1103/PhysRevLett.117.143602}
  {\bibfield  {journal} {\bibinfo  {journal} {Phys. Rev. Lett.}\ }\textbf
  {\bibinfo {volume} {117}},\ \bibinfo {pages} {143602} (\bibinfo {year}
  {2016})}\BibitemShut {NoStop}%
\bibitem [{\citenamefont {Nandi}\ \emph {et~al.}(2021)\citenamefont {Nandi},
  \citenamefont {An},\ and\ \citenamefont {Hosseini}}]{Nandi:21}%
  \BibitemOpen
  \bibfield  {author} {\bibinfo {author} {\bibfnamefont {A.}~\bibnamefont
  {Nandi}}, \bibinfo {author} {\bibfnamefont {H.}~\bibnamefont {An}},\ and\
  \bibinfo {author} {\bibfnamefont {M.}~\bibnamefont {Hosseini}},\ }\bibfield
  {title} {\bibinfo {title} {Coherent atomic mirror formed by randomly
  distributed ions inside a crystal},\ }\href
  {https://doi.org/10.1364/OL.423092} {\bibfield  {journal} {\bibinfo
  {journal} {Opt. Lett.}\ }\textbf {\bibinfo {volume} {46}},\ \bibinfo {pages}
  {1880} (\bibinfo {year} {2021})}\BibitemShut {NoStop}%
\bibitem [{\citenamefont {Pak}\ \emph {et~al.}(2022)\citenamefont {Pak},
  \citenamefont {Nandi}, \citenamefont {Titze}, \citenamefont {Bielejec},
  \citenamefont {Alaeian},\ and\ \citenamefont {Hosseini}}]{pak2022long}%
  \BibitemOpen
  \bibfield  {author} {\bibinfo {author} {\bibfnamefont {D.}~\bibnamefont
  {Pak}}, \bibinfo {author} {\bibfnamefont {A.}~\bibnamefont {Nandi}}, \bibinfo
  {author} {\bibfnamefont {M.}~\bibnamefont {Titze}}, \bibinfo {author}
  {\bibfnamefont {E.~S.}\ \bibnamefont {Bielejec}}, \bibinfo {author}
  {\bibfnamefont {H.}~\bibnamefont {Alaeian}},\ and\ \bibinfo {author}
  {\bibfnamefont {M.}~\bibnamefont {Hosseini}},\ }\bibfield  {title} {\bibinfo
  {title} {Long-range cooperative resonances in rare-earth ion arrays inside
  photonic resonators},\ }\href {https://doi.org/10.1038/s42005-022-00871-w}
  {\bibfield  {journal} {\bibinfo  {journal} {Comm. Phys.}\ }\textbf {\bibinfo
  {volume} {5}},\ \bibinfo {pages} {89} (\bibinfo {year} {2022})}\BibitemShut
  {NoStop}%
\bibitem [{\citenamefont {Jia}\ \emph {et~al.}(2022)\citenamefont {Jia},
  \citenamefont {Wu}, \citenamefont {Sun}, \citenamefont {Yan}, \citenamefont
  {Xie}, \citenamefont {Wang}, \citenamefont {Chen},\ and\ \citenamefont
  {Chen}}]{integrated_22}%
  \BibitemOpen
  \bibfield  {author} {\bibinfo {author} {\bibfnamefont {Y.}~\bibnamefont
  {Jia}}, \bibinfo {author} {\bibfnamefont {J.}~\bibnamefont {Wu}}, \bibinfo
  {author} {\bibfnamefont {X.}~\bibnamefont {Sun}}, \bibinfo {author}
  {\bibfnamefont {X.}~\bibnamefont {Yan}}, \bibinfo {author} {\bibfnamefont
  {R.}~\bibnamefont {Xie}}, \bibinfo {author} {\bibfnamefont {L.}~\bibnamefont
  {Wang}}, \bibinfo {author} {\bibfnamefont {Y.}~\bibnamefont {Chen}},\ and\
  \bibinfo {author} {\bibfnamefont {F.}~\bibnamefont {Chen}},\ }\bibfield
  {title} {\bibinfo {title} {Integrated photonics based on rare-earth ion-doped
  thin-film lithium niobate},\ }\href
  {https://doi.org/https://doi.org/10.1002/lpor.202200059} {\bibfield
  {journal} {\bibinfo  {journal} {Laser \& Photonics Reviews}\ }\textbf
  {\bibinfo {volume} {16}},\ \bibinfo {pages} {2200059} (\bibinfo {year}
  {2022})}\BibitemShut {NoStop}%
\bibitem [{\citenamefont {Kim}\ \emph {et~al.}(2021)\citenamefont {Kim},
  \citenamefont {Zhang}, \citenamefont {Ferreira}, \citenamefont {Banker},
  \citenamefont {Iverson}, \citenamefont {Sipahigil}, \citenamefont {Bello},
  \citenamefont {Gonz{\'a}lez-Tudela}, \citenamefont {Mirhosseini},\ and\
  \citenamefont {Painter}}]{kim2021quantum}%
  \BibitemOpen
  \bibfield  {author} {\bibinfo {author} {\bibfnamefont {E.}~\bibnamefont
  {Kim}}, \bibinfo {author} {\bibfnamefont {X.}~\bibnamefont {Zhang}}, \bibinfo
  {author} {\bibfnamefont {V.~S.}\ \bibnamefont {Ferreira}}, \bibinfo {author}
  {\bibfnamefont {J.}~\bibnamefont {Banker}}, \bibinfo {author} {\bibfnamefont
  {J.~K.}\ \bibnamefont {Iverson}}, \bibinfo {author} {\bibfnamefont
  {A.}~\bibnamefont {Sipahigil}}, \bibinfo {author} {\bibfnamefont
  {M.}~\bibnamefont {Bello}}, \bibinfo {author} {\bibfnamefont
  {A.}~\bibnamefont {Gonz{\'a}lez-Tudela}}, \bibinfo {author} {\bibfnamefont
  {M.}~\bibnamefont {Mirhosseini}},\ and\ \bibinfo {author} {\bibfnamefont
  {O.}~\bibnamefont {Painter}},\ }\bibfield  {title} {\bibinfo {title} {Quantum
  electrodynamics in a topological waveguide},\ }\href
  {https://doi.org/10.1103/physrevx.11.011015} {\bibfield  {journal} {\bibinfo
  {journal} {Phys. Rev. X}\ }\textbf {\bibinfo {volume} {11}},\ \bibinfo
  {pages} {011015} (\bibinfo {year} {2021})}\BibitemShut {NoStop}%
\bibitem [{\citenamefont {Mirhosseini}\ \emph {et~al.}(2019)\citenamefont
  {Mirhosseini}, \citenamefont {Kim}, \citenamefont {Zhang}, \citenamefont
  {Sipahigil}, \citenamefont {Dieterle}, \citenamefont {Keller}, \citenamefont
  {Asenjo-Garcia}, \citenamefont {Chang},\ and\ \citenamefont
  {Painter}}]{mirhosseini2019cavity}%
  \BibitemOpen
  \bibfield  {author} {\bibinfo {author} {\bibfnamefont {M.}~\bibnamefont
  {Mirhosseini}}, \bibinfo {author} {\bibfnamefont {E.}~\bibnamefont {Kim}},
  \bibinfo {author} {\bibfnamefont {X.}~\bibnamefont {Zhang}}, \bibinfo
  {author} {\bibfnamefont {A.}~\bibnamefont {Sipahigil}}, \bibinfo {author}
  {\bibfnamefont {P.~B.}\ \bibnamefont {Dieterle}}, \bibinfo {author}
  {\bibfnamefont {A.~J.}\ \bibnamefont {Keller}}, \bibinfo {author}
  {\bibfnamefont {A.}~\bibnamefont {Asenjo-Garcia}}, \bibinfo {author}
  {\bibfnamefont {D.~E.}\ \bibnamefont {Chang}},\ and\ \bibinfo {author}
  {\bibfnamefont {O.}~\bibnamefont {Painter}},\ }\bibfield  {title} {\bibinfo
  {title} {Cavity quantum electrodynamics with atom-like mirrors},\ }\href
  {https://doi.org/10.1038/s41586-019-1196-1} {\bibfield  {journal} {\bibinfo
  {journal} {Nature}\ }\textbf {\bibinfo {volume} {569}},\ \bibinfo {pages}
  {692} (\bibinfo {year} {2019})}\BibitemShut {NoStop}%
\bibitem [{\citenamefont {Douglas}\ \emph {et~al.}(2015)\citenamefont
  {Douglas}, \citenamefont {Habibian}, \citenamefont {Hung}, \citenamefont
  {Gorshkov}, \citenamefont {Kimble},\ and\ \citenamefont
  {Chang}}]{douglas2015quantum}%
  \BibitemOpen
  \bibfield  {author} {\bibinfo {author} {\bibfnamefont {J.~S.}\ \bibnamefont
  {Douglas}}, \bibinfo {author} {\bibfnamefont {H.}~\bibnamefont {Habibian}},
  \bibinfo {author} {\bibfnamefont {C.-L.}\ \bibnamefont {Hung}}, \bibinfo
  {author} {\bibfnamefont {A.~V.}\ \bibnamefont {Gorshkov}}, \bibinfo {author}
  {\bibfnamefont {H.~J.}\ \bibnamefont {Kimble}},\ and\ \bibinfo {author}
  {\bibfnamefont {D.~E.}\ \bibnamefont {Chang}},\ }\bibfield  {title} {\bibinfo
  {title} {Quantum many-body models with cold atoms coupled to photonic
  crystals},\ }\href {https://doi.org/10.1038/nphoton.2015.57} {\bibfield
  {journal} {\bibinfo  {journal} {Nat. Phot.}\ }\textbf {\bibinfo {volume}
  {9}},\ \bibinfo {pages} {326} (\bibinfo {year} {2015})}\BibitemShut {NoStop}%
\bibitem [{\citenamefont {See}\ \emph {et~al.}(2019)\citenamefont {See},
  \citenamefont {Bastidas}, \citenamefont {Tangpanitanon},\ and\ \citenamefont
  {Angelakis}}]{See2019}%
  \BibitemOpen
  \bibfield  {author} {\bibinfo {author} {\bibfnamefont {T.~F.}\ \bibnamefont
  {See}}, \bibinfo {author} {\bibfnamefont {V.~M.}\ \bibnamefont {Bastidas}},
  \bibinfo {author} {\bibfnamefont {J.}~\bibnamefont {Tangpanitanon}},\ and\
  \bibinfo {author} {\bibfnamefont {D.~G.}\ \bibnamefont {Angelakis}},\
  }\bibfield  {title} {\bibinfo {title} {Strongly correlated photon transport
  in nonlinear photonic lattices with disorder: Probing signatures of the
  localization transition},\ }\href
  {https://doi.org/10.1103/PhysRevA.99.033835} {\bibfield  {journal} {\bibinfo
  {journal} {Phys. Rev. A}\ }\textbf {\bibinfo {volume} {99}},\ \bibinfo
  {pages} {033835} (\bibinfo {year} {2019})}\BibitemShut {NoStop}%
\bibitem [{\citenamefont {Gonz{\'a}lez-Tudela}\ \emph
  {et~al.}(2017)\citenamefont {Gonz{\'a}lez-Tudela}, \citenamefont {Paulisch},
  \citenamefont {Kimble},\ and\ \citenamefont {Cirac}}]{gonzalez2017efficient}%
  \BibitemOpen
  \bibfield  {author} {\bibinfo {author} {\bibfnamefont {A.}~\bibnamefont
  {Gonz{\'a}lez-Tudela}}, \bibinfo {author} {\bibfnamefont {V.}~\bibnamefont
  {Paulisch}}, \bibinfo {author} {\bibfnamefont {H.}~\bibnamefont {Kimble}},\
  and\ \bibinfo {author} {\bibfnamefont {J.~I.}\ \bibnamefont {Cirac}},\
  }\bibfield  {title} {\bibinfo {title} {Efficient multiphoton generation in
  waveguide quantum electrodynamics},\ }\href
  {https://doi.org/10.1103/physrevlett.118.213601} {\bibfield  {journal}
  {\bibinfo  {journal} {Phys. Rev. Lett.}\ }\textbf {\bibinfo {volume} {118}},\
  \bibinfo {pages} {213601} (\bibinfo {year} {2017})}\BibitemShut {NoStop}%
\bibitem [{\citenamefont {Manassah}(2012)}]{Manassah_geometries}%
  \BibitemOpen
  \bibfield  {author} {\bibinfo {author} {\bibfnamefont {J.~T.}\ \bibnamefont
  {Manassah}},\ }\bibfield  {title} {\bibinfo {title} {Cooperative radiation
  from atoms in different geometries: decay rate and frequency shift},\ }\href
  {https://doi.org/10.1364/AOP.4.000108} {\bibfield  {journal} {\bibinfo
  {journal} {Adv. Opt. Photon.}\ }\textbf {\bibinfo {volume} {4}},\ \bibinfo
  {pages} {108} (\bibinfo {year} {2012})}\BibitemShut {NoStop}%
\bibitem [{\citenamefont {Braggio}\ \emph {et~al.}(2020)\citenamefont
  {Braggio}, \citenamefont {Chiossi}, \citenamefont {Carugno}, \citenamefont
  {Ortolan},\ and\ \citenamefont {Ruoso}}]{Braggio2020}%
  \BibitemOpen
  \bibfield  {author} {\bibinfo {author} {\bibfnamefont {C.}~\bibnamefont
  {Braggio}}, \bibinfo {author} {\bibfnamefont {F.}~\bibnamefont {Chiossi}},
  \bibinfo {author} {\bibfnamefont {G.}~\bibnamefont {Carugno}}, \bibinfo
  {author} {\bibfnamefont {A.}~\bibnamefont {Ortolan}},\ and\ \bibinfo {author}
  {\bibfnamefont {G.}~\bibnamefont {Ruoso}},\ }\bibfield  {title} {\bibinfo
  {title} {Spontaneous formation of a macroscopically extended coherent
  state},\ }\href {https://doi.org/10.1103/physrevresearch.2.033059} {\bibfield
   {journal} {\bibinfo  {journal} {Phys. Rev. Res.}\ }\textbf {\bibinfo
  {volume} {2}} (\bibinfo {year} {2020})}\BibitemShut {NoStop}%
\bibitem [{\citenamefont {Chang}\ \emph {et~al.}(2012)\citenamefont {Chang},
  \citenamefont {Jiang}, \citenamefont {Gorshkov},\ and\ \citenamefont
  {Kimble}}]{chang2012cavity}%
  \BibitemOpen
  \bibfield  {author} {\bibinfo {author} {\bibfnamefont {D.~E.}\ \bibnamefont
  {Chang}}, \bibinfo {author} {\bibfnamefont {L.}~\bibnamefont {Jiang}},
  \bibinfo {author} {\bibfnamefont {A.~V.}\ \bibnamefont {Gorshkov}},\ and\
  \bibinfo {author} {\bibfnamefont {H.~J.}\ \bibnamefont {Kimble}},\ }\bibfield
   {title} {\bibinfo {title} {Cavity {QED} with atomic mirrors},\ }\href
  {https://doi.org/10.1088/1367-2630/14/6/063003} {\bibfield  {journal}
  {\bibinfo  {journal} {New J. Phys.}\ }\textbf {\bibinfo {volume} {14}},\
  \bibinfo {pages} {063003} (\bibinfo {year} {2012})}\BibitemShut {NoStop}%
\bibitem [{\citenamefont {Moerner}\ and\ \citenamefont
  {Bjorklund}(1988)}]{moerner1988persistent}%
  \BibitemOpen
  \bibfield  {author} {\bibinfo {author} {\bibfnamefont {W.~E.}\ \bibnamefont
  {Moerner}}\ and\ \bibinfo {author} {\bibfnamefont {G.~C.}\ \bibnamefont
  {Bjorklund}},\ }\href {https://doi.org/10.1007/978-3-642-83290-1} {\emph
  {\bibinfo {title} {Persistent spectral hole-burning: science and
  applications}}},\ Vol.~\bibinfo {volume} {1}\ (\bibinfo  {publisher}
  {Springer},\ \bibinfo {year} {1988})\BibitemShut {NoStop}%
\bibitem [{\citenamefont {Ritter}\ \emph {et~al.}(2018)\citenamefont {Ritter},
  \citenamefont {Gruhler}, \citenamefont {Dobbertin}, \citenamefont
  {K{\"u}bler}, \citenamefont {Scheel}, \citenamefont {Pernice}, \citenamefont
  {Pfau},\ and\ \citenamefont {L{\"o}w}}]{ritter2018coupling}%
  \BibitemOpen
  \bibfield  {author} {\bibinfo {author} {\bibfnamefont {R.}~\bibnamefont
  {Ritter}}, \bibinfo {author} {\bibfnamefont {N.}~\bibnamefont {Gruhler}},
  \bibinfo {author} {\bibfnamefont {H.}~\bibnamefont {Dobbertin}}, \bibinfo
  {author} {\bibfnamefont {H.}~\bibnamefont {K{\"u}bler}}, \bibinfo {author}
  {\bibfnamefont {S.}~\bibnamefont {Scheel}}, \bibinfo {author} {\bibfnamefont
  {W.}~\bibnamefont {Pernice}}, \bibinfo {author} {\bibfnamefont
  {T.}~\bibnamefont {Pfau}},\ and\ \bibinfo {author} {\bibfnamefont
  {R.}~\bibnamefont {L{\"o}w}},\ }\bibfield  {title} {\bibinfo {title}
  {Coupling thermal atomic vapor to slot waveguides},\ }\href
  {https://doi.org/10.1103/physrevx.8.021032} {\bibfield  {journal} {\bibinfo
  {journal} {Phys. Rev. X}\ }\textbf {\bibinfo {volume} {8}},\ \bibinfo {pages}
  {021032} (\bibinfo {year} {2018})}\BibitemShut {NoStop}%
\bibitem [{\citenamefont {Norman}\ \emph {et~al.}(2019)\citenamefont {Norman},
  \citenamefont {Jung}, \citenamefont {Zhang}, \citenamefont {Wan},
  \citenamefont {Liu}, \citenamefont {Shang}, \citenamefont {Herrick},
  \citenamefont {Chow}, \citenamefont {Gossard},\ and\ \citenamefont
  {Bowers}}]{norman2019review}%
  \BibitemOpen
  \bibfield  {author} {\bibinfo {author} {\bibfnamefont {J.~C.}\ \bibnamefont
  {Norman}}, \bibinfo {author} {\bibfnamefont {D.}~\bibnamefont {Jung}},
  \bibinfo {author} {\bibfnamefont {Z.}~\bibnamefont {Zhang}}, \bibinfo
  {author} {\bibfnamefont {Y.}~\bibnamefont {Wan}}, \bibinfo {author}
  {\bibfnamefont {S.}~\bibnamefont {Liu}}, \bibinfo {author} {\bibfnamefont
  {C.}~\bibnamefont {Shang}}, \bibinfo {author} {\bibfnamefont {R.~W.}\
  \bibnamefont {Herrick}}, \bibinfo {author} {\bibfnamefont {W.~W.}\
  \bibnamefont {Chow}}, \bibinfo {author} {\bibfnamefont {A.~C.}\ \bibnamefont
  {Gossard}},\ and\ \bibinfo {author} {\bibfnamefont {J.~E.}\ \bibnamefont
  {Bowers}},\ }\bibfield  {title} {\bibinfo {title} {A review of
  high-performance quantum dot lasers on silicon},\ }\href
  {https://doi.org/10.1109/jqe.2019.2901508} {\bibfield  {journal} {\bibinfo
  {journal} {IEEE J. Quantum Electron.}\ }\textbf {\bibinfo {volume} {55}},\
  \bibinfo {pages} {1} (\bibinfo {year} {2019})}\BibitemShut {NoStop}%
\bibitem [{Note1()}]{Note1}%
  \BibitemOpen
  \bibinfo {note} {$\Gamma '$ may include photonic decay into unguided modes,
  or non-radiative decay into, e.g., material phonons.}\BibitemShut {Stop}%
\bibitem [{\citenamefont {Diniz}\ \emph {et~al.}(2011)\citenamefont {Diniz},
  \citenamefont {Portolan}, \citenamefont {Ferreira}, \citenamefont
  {G{\'e}rard}, \citenamefont {Bertet},\ and\ \citenamefont
  {Auffeves}}]{diniz2011strongly}%
  \BibitemOpen
  \bibfield  {author} {\bibinfo {author} {\bibfnamefont {I.}~\bibnamefont
  {Diniz}}, \bibinfo {author} {\bibfnamefont {S.}~\bibnamefont {Portolan}},
  \bibinfo {author} {\bibfnamefont {R.}~\bibnamefont {Ferreira}}, \bibinfo
  {author} {\bibfnamefont {J.}~\bibnamefont {G{\'e}rard}}, \bibinfo {author}
  {\bibfnamefont {P.}~\bibnamefont {Bertet}},\ and\ \bibinfo {author}
  {\bibfnamefont {A.}~\bibnamefont {Auffeves}},\ }\bibfield  {title} {\bibinfo
  {title} {Strongly coupling a cavity to inhomogeneous ensembles of emitters:
  Potential for long-lived solid-state quantum memories},\ }\href
  {https://doi.org/10.1103/physreva.84.063810} {\bibfield  {journal} {\bibinfo
  {journal} {Phys. Rev. A}\ }\textbf {\bibinfo {volume} {84}},\ \bibinfo
  {pages} {063810} (\bibinfo {year} {2011})}\BibitemShut {NoStop}%
\bibitem [{\citenamefont {Wang}\ \emph {et~al.}(2002)\citenamefont {Wang},
  \citenamefont {Wang}, \citenamefont {Gu},\ and\ \citenamefont
  {Yang}}]{wang2002}%
  \BibitemOpen
  \bibfield  {author} {\bibinfo {author} {\bibfnamefont {X.-H.}\ \bibnamefont
  {Wang}}, \bibinfo {author} {\bibfnamefont {R.}~\bibnamefont {Wang}}, \bibinfo
  {author} {\bibfnamefont {B.-Y.}\ \bibnamefont {Gu}},\ and\ \bibinfo {author}
  {\bibfnamefont {G.-Z.}\ \bibnamefont {Yang}},\ }\bibfield  {title} {\bibinfo
  {title} {Decay distribution of spontaneous emission from an assembly of atoms
  in photonic crystals with pseudogaps},\ }\href
  {https://doi.org/10.1103/PhysRevLett.88.093902} {\bibfield  {journal}
  {\bibinfo  {journal} {Phys. Rev. Lett.}\ }\textbf {\bibinfo {volume} {88}},\
  \bibinfo {pages} {093902} (\bibinfo {year} {2002})}\BibitemShut {NoStop}%
\bibitem [{\citenamefont {Asenjo-Garcia}\ \emph {et~al.}(2017)\citenamefont
  {Asenjo-Garcia}, \citenamefont {Hood}, \citenamefont {Chang},\ and\
  \citenamefont {Kimble}}]{asenjo2017atom}%
  \BibitemOpen
  \bibfield  {author} {\bibinfo {author} {\bibfnamefont {A.}~\bibnamefont
  {Asenjo-Garcia}}, \bibinfo {author} {\bibfnamefont {J.}~\bibnamefont {Hood}},
  \bibinfo {author} {\bibfnamefont {D.}~\bibnamefont {Chang}},\ and\ \bibinfo
  {author} {\bibfnamefont {H.}~\bibnamefont {Kimble}},\ }\bibfield  {title}
  {\bibinfo {title} {Atom-light interactions in quasi-one-dimensional
  nanostructures: {A} {G}reen's-function perspective},\ }\href
  {https://doi.org/10.1103/physreva.95.033818} {\bibfield  {journal} {\bibinfo
  {journal} {Phys. Rev. A}\ }\textbf {\bibinfo {volume} {95}},\ \bibinfo
  {pages} {033818} (\bibinfo {year} {2017})}\BibitemShut {NoStop}%
\bibitem [{\citenamefont {Liao}\ \emph {et~al.}(2016)\citenamefont {Liao},
  \citenamefont {Zeng}, \citenamefont {Nha},\ and\ \citenamefont
  {Zubairy}}]{liao2016photon}%
  \BibitemOpen
  \bibfield  {author} {\bibinfo {author} {\bibfnamefont {Z.}~\bibnamefont
  {Liao}}, \bibinfo {author} {\bibfnamefont {X.}~\bibnamefont {Zeng}}, \bibinfo
  {author} {\bibfnamefont {H.}~\bibnamefont {Nha}},\ and\ \bibinfo {author}
  {\bibfnamefont {M.~S.}\ \bibnamefont {Zubairy}},\ }\bibfield  {title}
  {\bibinfo {title} {Photon transport in a one-dimensional nanophotonic
  waveguide qed system},\ }\href
  {https://doi.org/10.1088/0031-8949/91/6/063004} {\bibfield  {journal}
  {\bibinfo  {journal} {Phys. Scripta}\ }\textbf {\bibinfo {volume} {91}},\
  \bibinfo {pages} {063004} (\bibinfo {year} {2016})}\BibitemShut {NoStop}%
\bibitem [{\citenamefont {Li}\ and\ \citenamefont {Wei}(2015)}]{Li2015}%
  \BibitemOpen
  \bibfield  {author} {\bibinfo {author} {\bibfnamefont {X.}~\bibnamefont
  {Li}}\ and\ \bibinfo {author} {\bibfnamefont {L.~F.}\ \bibnamefont {Wei}},\
  }\bibfield  {title} {\bibinfo {title} {Designable single-photon quantum
  routings with atomic mirrors},\ }\href
  {https://doi.org/10.1103/physreva.92.063836} {\bibfield  {journal} {\bibinfo
  {journal} {Phys. Rev. A}\ }\textbf {\bibinfo {volume} {92}} (\bibinfo {year}
  {2015})}\BibitemShut {NoStop}%
\bibitem [{\citenamefont {Domokos}\ \emph {et~al.}(2002)\citenamefont
  {Domokos}, \citenamefont {Horak},\ and\ \citenamefont
  {Ritsch}}]{Domokos2002}%
  \BibitemOpen
  \bibfield  {author} {\bibinfo {author} {\bibfnamefont {P.}~\bibnamefont
  {Domokos}}, \bibinfo {author} {\bibfnamefont {P.}~\bibnamefont {Horak}},\
  and\ \bibinfo {author} {\bibfnamefont {H.}~\bibnamefont {Ritsch}},\
  }\bibfield  {title} {\bibinfo {title} {Quantum description of light-pulse
  scattering on a single atom in waveguides},\ }\href
  {https://doi.org/10.1103/physreva.65.033832} {\bibfield  {journal} {\bibinfo
  {journal} {Physical Review A}\ }\textbf {\bibinfo {volume} {65}} (\bibinfo
  {year} {2002})}\BibitemShut {NoStop}%
\bibitem [{\citenamefont {Meng}\ \emph {et~al.}(2018)\citenamefont {Meng},
  \citenamefont {Dareau}, \citenamefont {Schneeweiss},\ and\ \citenamefont
  {Rauschenbeutel}}]{meng2018near}%
  \BibitemOpen
  \bibfield  {author} {\bibinfo {author} {\bibfnamefont {Y.}~\bibnamefont
  {Meng}}, \bibinfo {author} {\bibfnamefont {A.}~\bibnamefont {Dareau}},
  \bibinfo {author} {\bibfnamefont {P.}~\bibnamefont {Schneeweiss}},\ and\
  \bibinfo {author} {\bibfnamefont {A.}~\bibnamefont {Rauschenbeutel}},\
  }\bibfield  {title} {\bibinfo {title} {Near-ground-state cooling of atoms
  optically trapped 300 nm away from a hot surface},\ }\href
  {https://doi.org/10.1103/physrevx.8.031054} {\bibfield  {journal} {\bibinfo
  {journal} {Phys. Rev. X}\ }\textbf {\bibinfo {volume} {8}},\ \bibinfo {pages}
  {031054} (\bibinfo {year} {2018})}\BibitemShut {NoStop}%
\bibitem [{\citenamefont {Pfeiffer}\ \emph {et~al.}(2014)\citenamefont
  {Pfeiffer}, \citenamefont {Lindfors}, \citenamefont {Zhang}, \citenamefont
  {Fenk}, \citenamefont {Phillipp}, \citenamefont {Atkinson}, \citenamefont
  {Rastelli}, \citenamefont {Schmidt}, \citenamefont {Giessen},\ and\
  \citenamefont {Lippitz}}]{pfeiffer2014eleven}%
  \BibitemOpen
  \bibfield  {author} {\bibinfo {author} {\bibfnamefont {M.}~\bibnamefont
  {Pfeiffer}}, \bibinfo {author} {\bibfnamefont {K.}~\bibnamefont {Lindfors}},
  \bibinfo {author} {\bibfnamefont {H.}~\bibnamefont {Zhang}}, \bibinfo
  {author} {\bibfnamefont {B.}~\bibnamefont {Fenk}}, \bibinfo {author}
  {\bibfnamefont {F.}~\bibnamefont {Phillipp}}, \bibinfo {author}
  {\bibfnamefont {P.}~\bibnamefont {Atkinson}}, \bibinfo {author}
  {\bibfnamefont {A.}~\bibnamefont {Rastelli}}, \bibinfo {author}
  {\bibfnamefont {O.~G.}\ \bibnamefont {Schmidt}}, \bibinfo {author}
  {\bibfnamefont {H.}~\bibnamefont {Giessen}},\ and\ \bibinfo {author}
  {\bibfnamefont {M.}~\bibnamefont {Lippitz}},\ }\bibfield  {title} {\bibinfo
  {title} {Eleven nanometer alignment precision of a plasmonic nanoantenna with
  a self-assembled gaas quantum dot},\ }\href
  {https://doi.org/10.1021/nl403730q} {\bibfield  {journal} {\bibinfo
  {journal} {Nano Lett.}\ }\textbf {\bibinfo {volume} {14}},\ \bibinfo {pages}
  {197} (\bibinfo {year} {2014})}\BibitemShut {NoStop}%
\bibitem [{\citenamefont {Zhang}\ and\ \citenamefont
  {M{\o}lmer}(2019)}]{zhang2019theory}%
  \BibitemOpen
  \bibfield  {author} {\bibinfo {author} {\bibfnamefont {Y.-X.}\ \bibnamefont
  {Zhang}}\ and\ \bibinfo {author} {\bibfnamefont {K.}~\bibnamefont
  {M{\o}lmer}},\ }\bibfield  {title} {\bibinfo {title} {Theory of subradiant
  states of a one-dimensional two-level atom chain},\ }\href
  {https://doi.org/10.1103/physrevlett.122.203605} {\bibfield  {journal}
  {\bibinfo  {journal} {Phys. Rev. Lett.}\ }\textbf {\bibinfo {volume} {122}},\
  \bibinfo {pages} {203605} (\bibinfo {year} {2019})}\BibitemShut {NoStop}%
\bibitem [{\citenamefont {Vladimirova}\ \emph {et~al.}(1998)\citenamefont
  {Vladimirova}, \citenamefont {Ivchenko},\ and\ \citenamefont
  {Kavokin}}]{Vladimirova1998}%
  \BibitemOpen
  \bibfield  {author} {\bibinfo {author} {\bibfnamefont {M.~R.}\ \bibnamefont
  {Vladimirova}}, \bibinfo {author} {\bibfnamefont {E.~L.}\ \bibnamefont
  {Ivchenko}},\ and\ \bibinfo {author} {\bibfnamefont {A.~V.}\ \bibnamefont
  {Kavokin}},\ }\bibfield  {title} {\bibinfo {title} {Exciton polaritons in
  long-period quantum-well structures},\ }\href
  {https://doi.org/10.1134/1.1187364} {\bibfield  {journal} {\bibinfo
  {journal} {Semiconductors}\ }\textbf {\bibinfo {volume} {32}},\ \bibinfo
  {pages} {90} (\bibinfo {year} {1998})}\BibitemShut {NoStop}%
\bibitem [{\citenamefont {Ohta}\ \emph {et~al.}(2024)\citenamefont {Ohta},
  \citenamefont {Lelu}, \citenamefont {Xu}, \citenamefont {Inaba},
  \citenamefont {Hitachi}, \citenamefont {Taniyasu}, \citenamefont {Sanada},
  \citenamefont {Ishizawa}, \citenamefont {Tawara}, \citenamefont {Oguri},
  \citenamefont {Yamaguchi},\ and\ \citenamefont {Okamoto}}]{ohta2024acoustic}%
  \BibitemOpen
  \bibfield  {author} {\bibinfo {author} {\bibfnamefont {R.}~\bibnamefont
  {Ohta}}, \bibinfo {author} {\bibfnamefont {G.}~\bibnamefont {Lelu}}, \bibinfo
  {author} {\bibfnamefont {X.}~\bibnamefont {Xu}}, \bibinfo {author}
  {\bibfnamefont {T.}~\bibnamefont {Inaba}}, \bibinfo {author} {\bibfnamefont
  {K.}~\bibnamefont {Hitachi}}, \bibinfo {author} {\bibfnamefont
  {Y.}~\bibnamefont {Taniyasu}}, \bibinfo {author} {\bibfnamefont
  {H.}~\bibnamefont {Sanada}}, \bibinfo {author} {\bibfnamefont
  {A.}~\bibnamefont {Ishizawa}}, \bibinfo {author} {\bibfnamefont
  {T.}~\bibnamefont {Tawara}}, \bibinfo {author} {\bibfnamefont
  {K.}~\bibnamefont {Oguri}}, \bibinfo {author} {\bibfnamefont
  {H.}~\bibnamefont {Yamaguchi}},\ and\ \bibinfo {author} {\bibfnamefont
  {H.}~\bibnamefont {Okamoto}},\ }\bibfield  {title} {\bibinfo {title}
  {Observation of acoustically induced dressed states of rare-earth ions},\
  }\href {https://doi.org/10.1103/PhysRevLett.132.036904} {\bibfield  {journal}
  {\bibinfo  {journal} {Phys. Rev. Lett.}\ }\textbf {\bibinfo {volume} {132}},\
  \bibinfo {pages} {036904} (\bibinfo {year} {2024})}\BibitemShut {NoStop}%
\bibitem [{\citenamefont {Xu}\ \emph {et~al.}(2021)\citenamefont {Xu},
  \citenamefont {Inaba}, \citenamefont {Tsuchizawa}, \citenamefont {Ishizawa},
  \citenamefont {Sanada}, \citenamefont {Tawara}, \citenamefont {Omi},
  \citenamefont {Oguri},\ and\ \citenamefont {Gotoh}}]{xu2021low}%
  \BibitemOpen
  \bibfield  {author} {\bibinfo {author} {\bibfnamefont {X.}~\bibnamefont
  {Xu}}, \bibinfo {author} {\bibfnamefont {T.}~\bibnamefont {Inaba}}, \bibinfo
  {author} {\bibfnamefont {T.}~\bibnamefont {Tsuchizawa}}, \bibinfo {author}
  {\bibfnamefont {A.}~\bibnamefont {Ishizawa}}, \bibinfo {author}
  {\bibfnamefont {H.}~\bibnamefont {Sanada}}, \bibinfo {author} {\bibfnamefont
  {T.}~\bibnamefont {Tawara}}, \bibinfo {author} {\bibfnamefont
  {H.}~\bibnamefont {Omi}}, \bibinfo {author} {\bibfnamefont {K.}~\bibnamefont
  {Oguri}},\ and\ \bibinfo {author} {\bibfnamefont {H.}~\bibnamefont {Gotoh}},\
  }\bibfield  {title} {\bibinfo {title} {Low-loss {E}rbium-incorporated
  rare-earth oxide waveguides on {S}i with bound states in the continuum and
  the large optical signal enhancement in them},\ }\href
  {https://doi.org/10.1364/oe.437868} {\bibfield  {journal} {\bibinfo
  {journal} {Opt. Exp.}\ }\textbf {\bibinfo {volume} {29}},\ \bibinfo {pages}
  {41132} (\bibinfo {year} {2021})}\BibitemShut {NoStop}%
\bibitem [{\citenamefont {Viasnoff-Schwoob}\ \emph {et~al.}(2005)\citenamefont
  {Viasnoff-Schwoob}, \citenamefont {Weisbuch}, \citenamefont {Benisty},
  \citenamefont {Olivier}, \citenamefont {Varoutsis}, \citenamefont
  {Robert-Philip}, \citenamefont {Houdr{\'e}},\ and\ \citenamefont
  {Smith}}]{viasnoff2005spontaneous}%
  \BibitemOpen
  \bibfield  {author} {\bibinfo {author} {\bibfnamefont {E.}~\bibnamefont
  {Viasnoff-Schwoob}}, \bibinfo {author} {\bibfnamefont {C.}~\bibnamefont
  {Weisbuch}}, \bibinfo {author} {\bibfnamefont {H.}~\bibnamefont {Benisty}},
  \bibinfo {author} {\bibfnamefont {S.}~\bibnamefont {Olivier}}, \bibinfo
  {author} {\bibfnamefont {S.}~\bibnamefont {Varoutsis}}, \bibinfo {author}
  {\bibfnamefont {I.}~\bibnamefont {Robert-Philip}}, \bibinfo {author}
  {\bibfnamefont {R.}~\bibnamefont {Houdr{\'e}}},\ and\ \bibinfo {author}
  {\bibfnamefont {C.}~\bibnamefont {Smith}},\ }\bibfield  {title} {\bibinfo
  {title} {Spontaneous emission enhancement of quantum dots in a photonic
  crystal wire},\ }\href {https://doi.org/10.1103/physrevlett.95.183901}
  {\bibfield  {journal} {\bibinfo  {journal} {Phys. Rev. Lett.}\ }\textbf
  {\bibinfo {volume} {95}},\ \bibinfo {pages} {183901} (\bibinfo {year}
  {2005})}\BibitemShut {NoStop}%
\bibitem [{\citenamefont {Rao}\ and\ \citenamefont
  {Hughes}(2007)}]{rao2007single}%
  \BibitemOpen
  \bibfield  {author} {\bibinfo {author} {\bibfnamefont {V.~M.}\ \bibnamefont
  {Rao}}\ and\ \bibinfo {author} {\bibfnamefont {S.}~\bibnamefont {Hughes}},\
  }\bibfield  {title} {\bibinfo {title} {Single quantum-dot {P}urcell factor
  and $\beta$ factor in a photonic crystal waveguide},\ }\href
  {https://doi.org/10.1103/physrevb.75.205437} {\bibfield  {journal} {\bibinfo
  {journal} {Phys. Rev. B}\ }\textbf {\bibinfo {volume} {75}},\ \bibinfo
  {pages} {205437} (\bibinfo {year} {2007})}\BibitemShut {NoStop}%
\bibitem [{\citenamefont {G\"{u}sken}\ \emph {et~al.}(2023)\citenamefont
  {G\"{u}sken}, \citenamefont {Fu}, \citenamefont {Zapf}, \citenamefont
  {Nielsen}, \citenamefont {Dichtl}, \citenamefont {R\"{o}der}, \citenamefont
  {Clark}, \citenamefont {Maier}, \citenamefont {Ronning},\ and\ \citenamefont
  {Oulton}}]{Gsken2023}%
  \BibitemOpen
  \bibfield  {author} {\bibinfo {author} {\bibfnamefont {N.~A.}\ \bibnamefont
  {G\"{u}sken}}, \bibinfo {author} {\bibfnamefont {M.}~\bibnamefont {Fu}},
  \bibinfo {author} {\bibfnamefont {M.}~\bibnamefont {Zapf}}, \bibinfo {author}
  {\bibfnamefont {M.~P.}\ \bibnamefont {Nielsen}}, \bibinfo {author}
  {\bibfnamefont {P.}~\bibnamefont {Dichtl}}, \bibinfo {author} {\bibfnamefont
  {R.}~\bibnamefont {R\"{o}der}}, \bibinfo {author} {\bibfnamefont {A.~S.}\
  \bibnamefont {Clark}}, \bibinfo {author} {\bibfnamefont {S.~A.}\ \bibnamefont
  {Maier}}, \bibinfo {author} {\bibfnamefont {C.}~\bibnamefont {Ronning}},\
  and\ \bibinfo {author} {\bibfnamefont {R.~F.}\ \bibnamefont {Oulton}},\
  }\bibfield  {title} {\bibinfo {title} {Emission enhancement of erbium in a
  reverse nanofocusing waveguide},\ }\href
  {https://doi.org/10.1038/s41467-023-38262-6} {\bibfield  {journal} {\bibinfo
  {journal} {Nature Communications}\ }\textbf {\bibinfo {volume} {14}}
  (\bibinfo {year} {2023})}\BibitemShut {NoStop}%
\bibitem [{\citenamefont {Rackauckas}\ and\ \citenamefont
  {Nie}(2017)}]{rackauckas2017differentialequations}%
  \BibitemOpen
  \bibfield  {author} {\bibinfo {author} {\bibfnamefont {C.}~\bibnamefont
  {Rackauckas}}\ and\ \bibinfo {author} {\bibfnamefont {Q.}~\bibnamefont
  {Nie}},\ }\bibfield  {title} {\bibinfo {title} {Differential{E}quations.jl--a
  performant and feature-rich ecosystem for solving differential equations in
  {J}ulia},\ }\href@noop {} {\bibfield  {journal} {\bibinfo  {journal} {Journal
  of Open Research Software}\ }\textbf {\bibinfo {volume} {5}} (\bibinfo {year}
  {2017})}\BibitemShut {NoStop}%
\bibitem [{\citenamefont {Marzban}\ \emph {et~al.}(2015)\citenamefont
  {Marzban}, \citenamefont {Bartholomew}, \citenamefont {Madden}, \citenamefont
  {Vu},\ and\ \citenamefont {Sellars}}]{Marzban}%
  \BibitemOpen
  \bibfield  {author} {\bibinfo {author} {\bibfnamefont {S.}~\bibnamefont
  {Marzban}}, \bibinfo {author} {\bibfnamefont {J.~G.}\ \bibnamefont
  {Bartholomew}}, \bibinfo {author} {\bibfnamefont {S.}~\bibnamefont {Madden}},
  \bibinfo {author} {\bibfnamefont {K.}~\bibnamefont {Vu}},\ and\ \bibinfo
  {author} {\bibfnamefont {M.~J.}\ \bibnamefont {Sellars}},\ }\bibfield
  {title} {\bibinfo {title} {Observation of photon echoes from evanescently
  coupled rare-earth ions in a planar waveguide},\ }\href
  {https://doi.org/10.1103/PhysRevLett.115.013601} {\bibfield  {journal}
  {\bibinfo  {journal} {Phys. Rev. Lett.}\ }\textbf {\bibinfo {volume} {115}},\
  \bibinfo {pages} {013601} (\bibinfo {year} {2015})}\BibitemShut {NoStop}%
\bibitem [{\citenamefont {Chiossi}\ \emph {et~al.}(2022)\citenamefont
  {Chiossi}, \citenamefont {Lafitte-Houssat}, \citenamefont {Xia},
  \citenamefont {Sardi}, \citenamefont {Zhang}, \citenamefont {Welinski},
  \citenamefont {Berger}, \citenamefont {Morvan}, \citenamefont {Foteinou},
  \citenamefont {Ferrier}, \citenamefont {Serrano}, \citenamefont {Kolesov},
  \citenamefont {Wrachtrup},\ and\ \citenamefont
  {Goldner}}]{goldner2022waveguidedependence}%
  \BibitemOpen
  \bibfield  {author} {\bibinfo {author} {\bibfnamefont {F.}~\bibnamefont
  {Chiossi}}, \bibinfo {author} {\bibfnamefont {E.}~\bibnamefont
  {Lafitte-Houssat}}, \bibinfo {author} {\bibfnamefont {K.}~\bibnamefont
  {Xia}}, \bibinfo {author} {\bibfnamefont {F.}~\bibnamefont {Sardi}}, \bibinfo
  {author} {\bibfnamefont {Z.}~\bibnamefont {Zhang}}, \bibinfo {author}
  {\bibfnamefont {S.}~\bibnamefont {Welinski}}, \bibinfo {author}
  {\bibfnamefont {P.}~\bibnamefont {Berger}}, \bibinfo {author} {\bibfnamefont
  {L.}~\bibnamefont {Morvan}}, \bibinfo {author} {\bibfnamefont
  {V.}~\bibnamefont {Foteinou}}, \bibinfo {author} {\bibfnamefont
  {A.}~\bibnamefont {Ferrier}}, \bibinfo {author} {\bibfnamefont
  {D.}~\bibnamefont {Serrano}}, \bibinfo {author} {\bibfnamefont
  {R.}~\bibnamefont {Kolesov}}, \bibinfo {author} {\bibfnamefont
  {J.}~\bibnamefont {Wrachtrup}},\ and\ \bibinfo {author} {\bibfnamefont
  {P.}~\bibnamefont {Goldner}},\ }\bibfield  {title} {\bibinfo {title} {Photon
  echo, spectral hole burning, and optically detected magnetic resonance in
  $^{171}\mathrm{Yb}^{3+}$:${\mathrm{linbo}}_{3}$ bulk crystal and
  waveguides},\ }\href {https://doi.org/10.1103/PhysRevB.105.184115} {\bibfield
   {journal} {\bibinfo  {journal} {Phys. Rev. B}\ }\textbf {\bibinfo {volume}
  {105}},\ \bibinfo {pages} {184115} (\bibinfo {year} {2022})}\BibitemShut
  {NoStop}%
\end{thebibliography}

\clearpage

\appendix
\section{{Consideration of transverse positions}}
\label{app:transverse}
In order to take into account freedom in transverse position we bins the emitters according to frequency, longitudinal position along the waveguide, \textit{and also} transverse position in the cross-section of the waveguide, so that emitter $j$ is now indexed via $j \to (p,q,r)$ for $m$ longitudinal positional bins, $n'$ frequency bins, and $l$ transverse positional bins indexed by $p,q,r$ respectively with $lmn' = N$ and $l,m,n' \gg 1$. Assuming a single, common dipole transition coupling equally to forwards and backwards travelling fields, each emitter $(p,q,r)$ then experiences a coupling strength to the single waveguided mode that is proportional to $\sqrt{\Gamma_{r}}$. Waveguide-mediated coupling element between two emitters is now proportional to $\sqrt{\Gamma_{r}\Gamma_{r'}}.$ The coupling of emitters to the waveguided driving will also depend on this waveguide coupling strength, and so we factor the coherent driving as $\Omega_{j} \to {\Omega}_{p,r} =  \frac{\sqrt{l}\sqrt{\Gamma_{r}}}{\sqrt{\Gamma_{\text{1D}}}} {\Omega}_{p}$. Here, we have identified $\Gamma_{\text{1D}} = \frac{1}{l}\sum_{r}\Gamma_{r}$ as the average of waveguide decay rates experienced by the emitters distributed in the transverse plane. Similarly to the main text, the huge emitter number suggests a decorrelation of bins to create now a 3D grid. We obtain the equations of motion for the spins,
\begin{align}
    \dot{\sigma}_{p,q,r}^{-} &= \mathrm{i}\Delta_{q}\sigma_{p,q,r}^{-} -\frac{\Gamma'}{2}\sigma_{p,q,r}^{-} -  \frac{\sqrt{\Gamma_r}}{2}\sum_{p',q',r'}G_{p,p'}\sqrt{\Gamma_{r'}}\sigma^{-}_{p'q'r'} \nonumber\\ 
    &+ i\sqrt{l}\frac{\sqrt{\Gamma_{r}}}{\sqrt{\Gamma_{\text{1D}}}}{\Omega}_{p},
\label{eq-app:linear-eq}.
\end{align}
where the typical assumption is made that the emitters' frequency distribution and individual decay rate do not significantly depend on the emitter position within the waveguide, i.e., that the bulk properties of the emitters can be largely preserved~\cite{Marzban}. The overall inhomogeneous line may, however, differ from that of bulk media~\cite{goldner2022waveguidedependence} and so we assume its effect to be included in the distribution $\Delta_{q}$.
We form the collective spin $\hat{\sigma}_{p,q} = \frac{1}{\sqrt{\sum_{r}\Gamma_{r}}}\sum_{r}\sqrt{\Gamma_{r}}\hat{\sigma}_{p,q,r}$ from all the spins in a given transverse plane with a fixed position-frequency index pair $(p,q).$ We then obtain the result analogous to the main text
\begin{align}
    \dot{\sigma}_{p,q}^{-} &= \mathrm{i}\Delta_{q}\sigma_{p,q}^{-} -\frac{\Gamma'}{2}\sigma_{p,q}^{-} -  \frac{l\Gamma_{\text{1D}}}{2}\sum_{p',q'}G_{p,p'}\sigma^{-}_{p'q'} \nonumber\\
    &+ i\sqrt{l}\Omega_{p},
\label{eq-app:linear-eq-2}
\end{align}
We can further obtain the linear-response relation upon forming the collective spin via a sum over $q$ as in the main text:
\begin{eqnarray}
\label{eq-app:linear_response}
    \mathcal{B}_{p}^{-} = i\sqrt{n'l}\gamma_{\text{inh}}^{-1}\chi(\Delta_{c})\left[ -  \frac{\sqrt{n'l}\Gamma_{\text{1D}}}{2}\sum_{p'=1}^{m}G_{p,p'}\mathcal{B}_{p'}^{-} + i\Omega_{p}\right],
\end{eqnarray}
which depends on the product $\sqrt{n'l}$. Setting $n = n'l$ then reproduces the result \eqref{eq:linear_response}. It is thus appropriate to assume in the single-excitation subspace that each `emitter' in the main text already accounts the effects of individual emitter variation in the transverse plane, keeping in mind that $\Gamma_{\text{1D}}$ represents an average of waveguide decay rates over the transverse plane. 
\section{Considerations for the binning procedure}
\label{app:binning} 

\begin{figure}[h!]
    \centering
    \includegraphics[width=\linewidth]{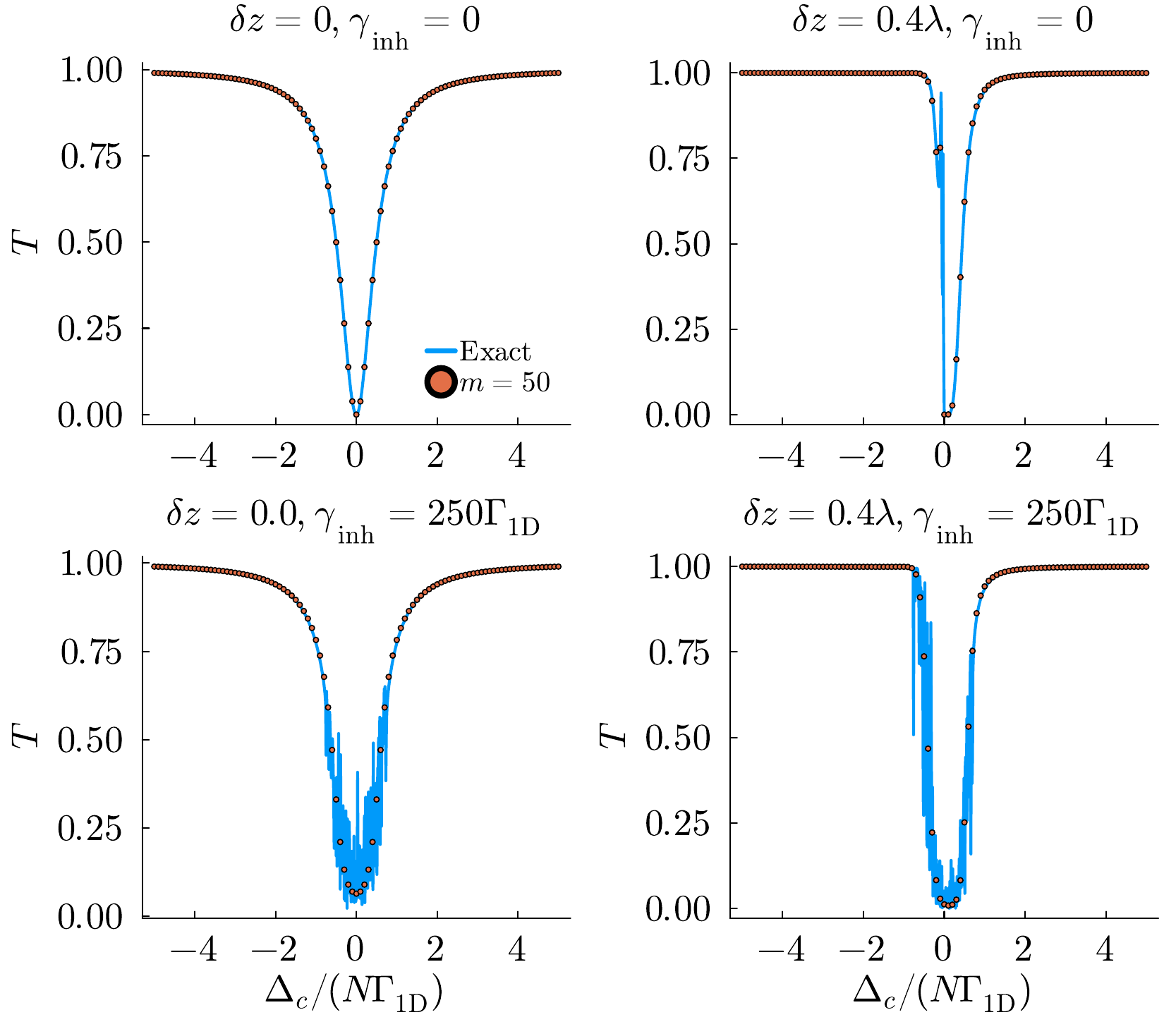}
    \caption{Transmission through an ensemble of size $N=1000$ using the exact expression~\cite{asenjo2017atom} (blue line), and using the binning procedure with $m=50$ (orange dots) for $\Gamma' = \Gamma_{\text{1D}}$ and varying $\gamma_{\text{inh}}, \delta z$. For the exact calculation we sample each emitter position $z_{j} \sim U(0,\delta z)$ and each detuning $\Delta_{j} \sim \text{N}(0,\gamma_{\text{inh}}/\sqrt{\text{ln}2})$ from the Gaussian distribution $\text{N}$ giving a FWHM $\gamma_{\text{inh}}.$ For the approximation, the binning procedure is applied according to the main text wiith $z_{p} \sim U(0,\delta z).$}
    \label{fig:app-verification}
\end{figure}

Given the $n\times m$ binning procedure of $N$ emitters, both $n$ and $m$ must be chosen large enough to allow for their respective continuum approximations to be valid so that spin dynamics of the original system of size $N$ may be well approximated by the dynamics of the $m$ non-Lorentzian collective spins. {In particular, we wish to well-approximate system dynamics over the fastest timescale $\sim \text{Min}((N\Gamma_{\text{1D}})^{-1},\gamma_{\text{inh}}^{-1}),$ and the $n\times m$ binning procedure can be expected to introduce discrepancies on the much slower timescales $O(n\gamma_{\text{inh}}^{-1})$ and $O(m(N\Gamma_{\text{inh}})^{-1})$ for $m, n \gg 1$. These correspond to discrepancies in the spectrum with widths orders of magnitude narrow than the broadest features typically of interest}. After the renormalization over frequency is additionally made, the resulting system matrix for steady-state dynamics is $m\times m$, which should be small enough to allow numerical computation of eigenvalues (i.e., $m\sim 10^4$ for a standard desktop). For positional fluctuations much smaller than a wavelength, $n=10^6$ and $m=10^3$ produce satisfactory results in the main text with small fluctuations over individual realisations [as in Fig.~\ref{fig:strong_coupling}(b)] and little variation as $n,m$ are locally varied with constant $nm = N$. The results are expected to remain satisfactory for smaller $N$ such that $n\gtrapprox 10^3.$ For smaller system sizes than this exact numerical computation should be then feasible. Whilst verification is not possible for $N\sim 10^9,$ good agreement between exact transmission through an ensemble of size $N=1000$ and the approximate system with $m=50$ bins is shown in Fig. \ref{fig:app-verification} for varying positional ensemble extent and Gaussian inhomogeneous broadening. Taking the continuum approximation $\chi$ of the response function precludes the presence of narrow spectral features of width $\sim \Gamma'$ in the approximate spectrum, but the broad features of practical interest are well-captured.
\section{Response functions}\label{app:response}
We here list the well-known results~\cite{diniz2011strongly} of (unnormalized) response functions 
$W_{s}(\Delta_{c}) = \int \frac{d \Delta'\rho_{s}(\Delta')}{\Delta_{c} - \Delta'  + \mathrm{i}\Gamma'/2}$ for Gaussian ($s = g$), Uniform ($s = u$), and Lorentzian ($s = l$) distributions with a FWHM $\gamma_{\text{inh}}$ respectively as follows:
\begin{align}
    W_{g}(\Delta_{c}) &= \frac{\sqrt{\pi}}{\mathrm{i}\gamma_{\text{inh}}/2}\frac{\sqrt{\text{ln}2}}{2}\text{erfcx}\left(\frac{\sqrt{\text{ln}2}}{2}\frac{\Delta_{c} + \mathrm{i}\Gamma'/2}{\mathrm{i}\gamma_{\text{inh}}/2}\right) \\ W_{u}(\Delta_{c}) &= \frac{1}{\mathrm{i}\gamma_{\text{inh}}/2}\text{arctan}\left(\frac{\mathrm{i}\gamma_{\text{inh}}/2}{ \Delta_{c} + \mathrm{i}\Gamma'/2}\right) \\ 
    W_{l}(\Delta_{c}) &= \frac{1}{\Delta_{c} + \mathrm{i}(\Gamma' + \gamma_{\text{inh}})/2},
\end{align}
where the densities read
\begin{align}
    \rho_{g}(\Delta_{c}) &= \frac{1}{(\gamma_{\text{inh}}/\sqrt{\text{ln}2})\sqrt{\pi}}e^{-\Delta_{c}^2/(\gamma_{\text{inh}}/\sqrt{\text{ln}2})^2} \\
    \rho_{u}(\Delta_{c}) &= \frac{\chi_{[-\gamma_{\text{inh}}/2,\gamma_{\text{inh}}/2]}}{\gamma_{\text{inh}}} \\
    \rho_{l}(\Delta_{c}) &= \frac{(\gamma_{\text{inh}}/2)}{\pi}\frac{1}{(\gamma_{\text{inh}}/2)^2 + \Delta_{c}^2},
\end{align}
with the indicator function $\chi_{[a,b]}$ taking value $1$ in $[a,b]$ and $0$ otherwise.

\section{{Eigenvalues via continuum limit}}\label{app:eigval_approx}
\begin{figure}
    \centering
    \vspace{20pt}
    \includegraphics[width=1.0\linewidth]{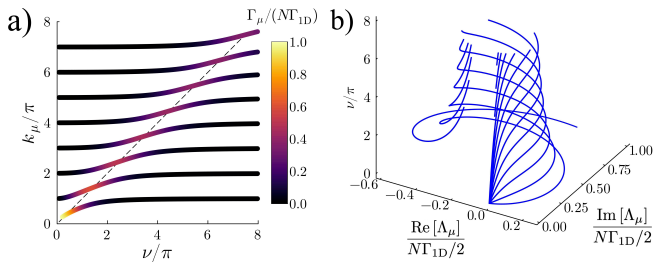}
    \caption{(a) First 8 wavevectors $k_{\mu}$ of the spin wave eigenmodes calculated using Eq. \eqref{eq:consistency_condtion} for spatially extended ensembles. The colorbar denotes decay rate of the collective mode (calculated using \eqref{eq:general-eigenvalue}) normalized to the collective decay rate of a point-like ensemble. The dashed line denotes $k_{\mu} = \nu$, and highlights that the eigenmode with maximal decay rate has a wavevector approximately lying on this line (i.e., the eigenmode wavelength is approximately the length of the ensemble sample). For large $\nu \gg 1$ the eigenmode with maximal decay observes a local maximum of its decay rate at $k_{\mu} \sim \nu \sim (n+1/2)\pi.$ However, this maximum is decreasing with increasing $\nu.$ (b) Normalized eigenvalues in the complex plane as a function of $\nu$. Away from $\nu \approx 0$, there is in general more than one eigenmode with appreciable (normalized) decay rate.}
    \label{fig:app:eigenvalues}
\end{figure}

To obtain analytical expressions for system eigenvalues we consider the distributions of atoms uniform on $[0,\delta z]$ and with equal spacing $\delta z/m$. At the expense of neglecting to treat states with smaller decay rates, the eigenvalues of the pseudo-random, equally spaced system can well reproduce behaviour on the shorter time scales associated with broad resonances of the collective spin even in the fully random system. This is due to the fact that the broadest resonances are associated with slowly varying polarization profiles across the ensemble, which are negligibly perturbed by the positional fluctuations on much shorter length scales. Assuming emitter density is high (i.e., many emitters in a given wavelength), we move to the continuum limit of the eigenvalue problem for $\Lambda_{\mu}$:

\begin{equation}
	\frac{\mathrm{i}N\Gamma_{\text{1D}}}{2}\int_{0}^{\delta z}\frac{dz'}{\delta z} \exp(\mathrm{i}\beta|z-z'|)\sigma_{\mu}(z') = \Lambda_{\mu} \sigma_{\mu}(z),
 \label{eq-app:eigenvalue-eq}
\end{equation}
which can be transformed to the unit interval:
\begin{equation}
    \frac{\mathrm{i}N\Gamma_{\text{1D}}}{2}\int_{0}^{1}dZ'\exp(\mathrm{i}\nu|Z-Z'|)\tilde{\sigma}_{\mu}(Z') = \Lambda_{\mu} \tilde{\sigma}_{\mu}(Z),
\end{equation}
where $Z = z/(\delta z)$ and ${\sigma}(z)$ is the spin profile at continuum position $z$, and $\tilde{\sigma}_{\mu}(Z) = \sigma_{\mu}(z)$. Whilst the uniform distribution is considered here, the corresponding expression for general distribution immediately shows that in the high density limit the existing eigenvalues of the system are unchanging up to a scaling with $m$, regardless of the underlying positional distribution. Similarly to the discrete case~\cite{Vladimirova1998},  the spin profile
\begin{equation}
	\tilde{\sigma}_{\mu}(Z) = \mathcal{N}_{\mu}\left(e^{\mathrm{i}k_{\mu}Z} + e^{\mathrm{i}\phi_{\mu}}e^{-\mathrm{i}k_{\mu}Z}\right),
\end{equation}
for $k_{\mu} \neq 0$ (when $\nu \neq 0)$ yields the eigenvalue 
\begin{equation}	\label{eq:general-eigenvalue}
\Lambda_{\mu} = \frac{\mathrm{i}N\Gamma_{\text{1D}}}{2}\frac{2\mathrm{i}\nu}{\nu^2 - k_{\mu}^2}
\end{equation}
subject to the transcendental equation for $k_{\mu}$
\begin{equation}
	\label{eq:consistency_condtion}
	\left(\frac{k_{\mu}+\nu}{k_{\mu}-\nu}\right)^2 = e^{2\mathrm{i}k_{\mu}} = e^{2\mathrm{i}\phi_{\mu}},
\end{equation}
for normalization factor $\mathcal{N}_{\mu}$, which holds for arbitrary ensemble extent in the high density limit. The limit $\nu = 0$ yields $k_{\mu} = \mu\pi, {\mu} = 0,1,\ldots$, corresponding to $m-1$ states with zero decay rate and the single broad resonance (for which a more careful limiting argument is required). Assuming $\nu \ll \pi,$ one may perturbatively solve \eqref{eq:consistency_condtion} for $\mu\geq 1$ to yield the approximate spin wavenumbers
\begin{equation}
	k_{\mu} = \mu \pi - \frac{2\mathrm{i}}{\mu\pi}\nu + \frac{4}{\mu^3\pi^3}\nu^2 + O(\nu^3),
\end{equation}
whilst for $\mu = 0$, $k_{\mu}$ scales as $\sqrt{\nu}$. One can carry out the perturbation analysis or just note the trace property $\sum_{\mu=0}^{\infty}\Lambda_{\mu} = \frac{\mathrm{i}N\Gamma_{\text{1D}}}{2}$. Inserting $k_{\mu}$ into the eigenvalue and using this relation (and standard identities for $\sum_{s=1}^{\infty}\frac{1}{s^{2}},$ etc.,) gives the eigenvalues in presented in the main text. Note that the discussion up to and including \eqref{eq:transmission} remains valid for even a spatially extended ensemble when using the general eigenvalue \eqref{eq:general-eigenvalue}, so that transmission through a spatially extended and spectrally inhomogeneous ensemble may also be in principle calculated numerically. For instance, the eigenvalues and wavevectors of the system \eqref{eq-app:eigenvalue-eq} are calculated in this way and presented in Fig. \ref{fig:app:eigenvalues}. Whilst a detailed analysis of extended ensembles lies outside of the scope of this work, we chiefly note that the limit $\nu \to 0$ maximizes the decay of a single (symmetric) excitation into the waveguide, whilst minimizing coupling to of all other collective excitations. Indeed, once $\nu$ becomes appreciably greater than zero then multiple eigenmodes will be in general appreciably excited by a waveguide driving. In this sense, spatially localized ensembles are preferred over extended ensembles when aiming to coherently interface with single collective excitations only for the purposes of, e.g., a cavity QED implementation. As such ensembles are additionally both simpler to address theoretically and relevant experimentally~\cite{Nandi:21,pak2022long} they are our focus in the main text.
\section{Perturbative coupling to narrower resonances}\label{app:dark_coupling}
In the limit $\nu \to 0,$ the spin profiles tend to $\tilde{\sigma}_{\mu}(Z) = (1/\mathcal{N}_{\mu})\cos(\mu\pi Z)$ for $\mu=0,1,\ldots.$ The coupling between the state with $\mu = 0$ and other states $\mu \neq 0$ via the non-Hermitian Hamiltonian is approximated at leading order by $\frac{\mathrm{i}\sqrt{2}mn\Gamma_{\text{1D}}}{2}\int_{[0,1]^2}dZdZ'\cos(\mu\pi Z)\exp(\mathrm{i}\nu|Z-Z'|) =   \frac{mn\Gamma_{\text{1D}}}{2\sqrt{2}\mu^2\pi^2}\nu,$ for even $\mu$ and 0 for odd $\mu.$ A rudimentary rate of loss of coherence in the $\mu = 0$ state (i.e., of $\mathfrak{B}^{-}$ for the single localized ensemble in the main text) is then given as the sum of these rates for $\mu=1,\ldots$. The sum is evaluated as $\frac{N\Gamma_{\text{1D}}}{12\sqrt{2}}\nu$, which yields the result of the main text upon replacement of $\nu$ and approximation $\pi/(3\sqrt{2}) \approx 1.$

\end{document}